\documentclass[twocolumn]{aastex63}
\usepackage{amsmath}
\usepackage{times}
\usepackage{graphicx}
\usepackage{xcolor}
\usepackage{bm}
\usepackage{hyperref}
\usepackage{multirow}
\hypersetup{colorlinks = true, linkcolor = blue, anchorcolor = red, 
     citecolor = blue, filecolor = red, urlcolor = blue, backref = page}

\usepackage[normalem]{ulem}

\begin{document}
\title{Infra-Red Emission from Cold Gas Dusty Disks in Massive Ellipticals}

\author{Zhaoming Gan}
\affiliation{Department of Astronomy, Columbia University, 550 W, 120th Street, New York, NY 10027, USA}

\author{Brandon S. Hensley}
\affiliation{Spitzer Fellow, Department of Astrophysical Sciences,  Princeton University, Princeton, NJ 08544, USA}

\author{Jeremiah P. Ostriker}
\affiliation{Department of Astronomy, Columbia University, 550 W, 120th Street, New York, NY 10027, USA}

\author{Luca Ciotti}
\affiliation{Department of Physics and Astronomy, University of Bologna, via Piero Gobetti 93/2, 40129 Bologna, Italy}

\author{David Schiminovich}
\affiliation{Department of Astronomy, Columbia University, 550 W, 120th Street, New York, NY 10027, USA}

\author{Silvia Pellegrini}
\affiliation{Department of Physics and Astronomy, University of Bologna, via Piero Gobetti 93/2, 40129 Bologna, Italy}

\begin{abstract}
What is the expected infrared output of elliptical galaxies? 
We report the latest findings obtained in this high time resolution 
($\sim10$ years) and high spatial resolution ($2.5$ parsec at center) study.
We add a set of grain physics to the \texttt{MACER} code, including
  (a) dust grains made in passive stellar evolution; 
  (b) dust grain growth due to collision and sticking; 
  (c) grain destruction due to thermal sputtering; 
  (d) dust cooling of hot gas via inelastic collisions; and 
  (e) radiation pressure on dust grains. 
The code improvements enable us to analyze metal depletion 
and AGN obscuration due to dust, and to assess its infrared output. 
We simulate a representative massive elliptical galaxy 
of a central stellar velocity dispersion $\sim260$ km/s 
and modest rotation. 
We find that:  
(1)	the circumnuclear disk (of a size {$\sim 1$ kpc}) is dusty 
        in its outer region where most of the metals are in dust grains, 
        while in the inner disk dust grains are mostly destroyed by the AGN irradiation; 
(2)	the dusty disk is optically thick to both the starlight within the disk 
        and the radiation from the central AGN. 
        Thus the AGN is obscured behind the disk, the covering factor is $\sim0.2$; 
(3)	the duty cycles of the AGN activities, star formation, 
        and the dust infrared luminosity roughly match observations, 
        e.g., in most of its lifetime, the simulated galaxy is 
        a stereotypical ``quiescent" elliptical galaxy 
        with {$L_{\rm IR}\sim 10^{11}L_\odot$}, 
        bu it can reach $\gtrsim10^{46}$ erg/s during outbursts 
        with the star formation rate {$\gtrsim250~M_\odot/{\rm yr}$}.
\end{abstract}

\keywords{black hole physics---
                 galaxies: elliptical and lenticular, cD---
                 galaxies: evolution---
                 ISM: abundances---
                 ISM: dust---
                 infrared: galaxies---
                 methods: numerical}

\section{Introduction}
 \label{sec:introduction}

The classic picture of massive elliptical galaxies imagines a quasi-spherical system filled with old, low mass stars 
and hot, thermal X-ray emitting, moderately metal rich gas. In fact there has been evidence known for decades 
that the central regions often contain significant amounts of cold gas, dust and young stars. 
As early as the mid 1970s, Knapp and associates authored several papers
on the central regions of normal elliptical galaxies with significant evidence for dust 
 (e.g., \citealt{knapp_study_1975, knapp_neutral_1978}).
The prevalence of central dusty disks was noted as far back as \citet{van_dokkum_dust_1995}.
Early guesses for the origin of the observed cold component include cooling flows and mergers. 
\textcolor{black}{However,} there has been very little work done in trying to understand and model these phenomena for two reasons.
First, the standard cosmological codes have a spatial resolution of roughly in the range $0.1-1.0$ kpc \citep{sijacki_illustris_2015,hopkins_fire-2_2018}, 
so they are not equipped to study phenomena occurring in the inner 50 parsec of these systems.
\textcolor{black}{Second,} the prevailing characterization of these systems as ``quiescent" 
made evidence for cold gas and star formation seem ``anomalous".

\textcolor{black}{
Our} previous work at high resolution (inner radius $\sim 2.5$ parsec)
has shown us that intermittent cooling flows are a normal part of the evolution of these systems.
Consequently, cool gas collects in central nuclear disks, 
\textcolor{black}{inducing both feeding of the central black holes and episodes of star formation characterized by a top-heavy initial mass function.}
In this paper, we turn our focus to the resultant infrared output and compare with observations.
We will see that the cold gas component can exceed $10^{10}M\odot$ at certain times with the dust mass
exceeding $10^8M_\odot$ in accoord with the ALMA/SCUBA-2 observations 
such as those by \citet{dudzeviciute_alma_2019} and \citet{lang_revealing_2019}.

\textcolor{black}{
In  this series of papers, our aim is to simulate and to understand the black hole feeding and AGN feedback 
processes in isolated massive elliptical galaxies. 
We also use the results from cosmological simulations to guide our implementation of cosmic accretion and its chemical properties. 
For example, \citet{choi_physics_2017} performed zoom-in cosmological simulations on individual galaxies to study
the effects of AGN feedback and the chemical evolution in the galaxies and their surroundings 
(see also \citealt{eisenreich_active_2017,brennan_momentum-driven_2018}).}
%
In our \texttt{MACER} simulations, a relatively complete set of stellar physics has been incorporated, 
including AGB stars, and supernovae of type I and II. 
We calculate the nucleosynthesis output from the stars and thereby 
track the chemical evolution in the modeled galaxies. 
\textcolor{black}{Following \citet{choi_physics_2017},} 
we add a suite of chemical abundances to the code, solving 12 additional continuity equations for
H, He, C, N, O, Ne, Mg, Si, S, Ca, Fe and Ni respectively 
with metal yields from dying stars based on standard stellar physics \citep{gan_adding_2019}. 
With the high resolution of a few parsecs in central regions, we can track the metal enrichment, 
transportation and dilution throughout the modeled galaxies.
As an example, we paid special attention to the chemical composition 
of the BAL (Broad Absorption Line) winds. We find that 
the simulated metallicity  in the BAL winds could be up to $\sim 8 Z_\odot$, 
matching well with SDSS observations of BLR gas
\citep{nagao_evolution_2006, xu_evolution_2018}.

In our previous \texttt{MACER} simulations, 
we have not corrected for the depletion of metals
onto dust, while the fractional depletion of refractory elements onto dust grains
may be a large correction in the cold gas component, e.g., in/around the circumnuclear disk
(see \citealt{hensley_grain_2014} for the case without rotation, nor metal tracers).
\textcolor{black}{Previous work on dust production in elliptical galaxies and its chemical effects 
can be also found in the papers by \citet{Calura_cycle_2008} (in the post-starburst phase), and 
\citet{Pipino_chemical_2011} and \citet{Gall_genesis_2011} (during the starburst phase). }
In this paper, we will discuss our implementation of grain physics into the \texttt{MACER} code,
and we will show how periodic outbursts of star formation and AGN activity 
can cause copious emission of IR radiation from the dust. The peak IR luminosity can reach $10^{46}$ erg/s.
We organize the rest of the paper as follows: For completeness, we recall briefly the hydrodynamical equations 
and the chemical tracers  in \S\ref{sec:hydro} and \S\ref{sec:metals}, respectively 
(see \citealt{gan_macer_2019, gan_adding_2019} for more details). 
In \S\ref{sec:grain-physics}, we introduce the grain physics proposed by \citet{hensley_grain_2014}, 
and describe its implementation in the \texttt{MACER} code. 
In \S\ref{sec:model-updates} we outline the additional physical processes that we have included. 
We focus on the observable infrared outputs in \S\ref{sec:results} 
and compare them with observations in \S\ref{sec:compare-to-obs}.
Finally, conclusions are presented in \S\ref{sec:conclusions}.

\section{Hydrodynamics} \label{sec:hydro}
We solve the time-dependent Eulerian equations which governs the hydrodynamics of 
the gaseous interstellar medium (ISM) over a length scale from 2.5 parsec (inner boundary) to 
250 kilo-parsec (outer boundary; 
\citealt{ciotti_cooling_1997, novak_feedback_2011, gan_macer_2019}), 
\begin{equation} \label{eq:massconsvr}
 \begin{aligned}
    \frac{\partial \rho_{\rm gas}}{\partial t} &+ \nabla\cdot(\,\rho_{\rm gas}\,{\bf v}\,) =  
                -\nabla\cdot{\bf T}   - \dot{\rho}_{\star}^{+} \quad\quad\quad\quad\quad\quad\quad\quad\quad\quad \\
           &  +\dot{\rho}_{\rm II}\cdot R_{\rm II, gas}+\dot{\rho}_{\rm I}\cdot R_{\rm I, gas} 
           +\dot{\rho}_{\star}\cdot R_{\rm \star, gas} 
  \end{aligned}
\end{equation}
\begin{equation} \label{eq:momconsvr}
 \begin{aligned}
   \frac{\partial {\bf m}}{\partial t} &+ \nabla\cdot({\bf m v}) =
           - \nabla p_{\rm gas} - \nabla p_{\rm rad} - \rho_{\rm gas}\nabla \phi   -\dot{\bf m}^{+}_{\star} \quad\\
        & - \nabla\cdot\Pi_{\rm vis} - \nabla\cdot\Pi_{\rm T} +\dot{\bf m}_{\rm S}
           + \rho_{\rm gas}\cdot {\bf f}_{\rm rad,dust},
  \end{aligned}
\end{equation}
\begin{equation} \label{eq:engconsvr}
 \begin{aligned}
   \frac{\partial E}{\partial t} &+ \nabla\cdot(E{\bf v}) =
            - p_{\rm gas} \nabla \cdot {\bf v}  + H - C + \dot{E}_{\rm T}  -\dot{E}^{+}_{\star} \\
            & 
            - \Pi_{\rm vis}:\nabla {\bf v} +\dot{E}_{\rm II}+\dot{E}_{\rm I} + \dot{E}_{\rm S} - H^{\rm gas}_{\rm dust,collision}, \quad\quad
  \end{aligned}
\end{equation}
where $\rho_{\rm gas}$, ${\bf m}$, $E$ are the mass, momentum 
and energy densities for the gaseous ISM, respectively. 
As usual, $p_{\rm gas}$ is the gas thermal pressure and ${\bf v}$ is its velocity. 
The adiabatic index is fixed to $\gamma=5/3$.
We refer the readers to our code paper (\citealt{gan_macer_2019}, and references therein) 
for the details of the physics we added in the simulation. 
We assemble the galaxy dynamics, stellar evolution and atomic physics 
into our hydrodynamical simulations in form of source/sink terms and boundary conditions,
which are outlined briefly as follows:

\begin{enumerate}
\item {\bf Galaxy dynamics} of rotating elliptical galaxies.
Detailed modeling of the galaxy dynamics, 
including the gravitational potential $\phi$ and the stellar mass distribution (of total mass $M_\star$), 
are taken as the ``background'' of the simulation (cf. \S\ref{sec:galaxy-model}). 
Rotation is allowed and determined self-consistently by solving the Jeans' equations. 
%
\item {\bf Stellar evolution.}
Passive stellar evolution is considered as mass and energy sources
\citep{ciotti_agn_2012, pellegrini_hot_2012}, including 
AGB stars  ($\dot{\rho}_{\star} $, $\dot{E}_{\rm S} $),
SNe Ia ($\dot{\rho}_{\rm I} $, $\dot{E}_{\rm I} $) (from the aged stellar population mentioned above),
and SNe II ($\dot{\rho}_{\rm II}$, $\dot{E}_{\rm II}$) (if any, due to star formation; see below),
where $R_{\rm \star, gas}$, $R_{\rm I, gas}$, and $R_{\rm II, gas}$ 
(see Equation \ref{eq:stellar-ejecta-dust-ratio})
are the mass fractions of the stellar ejecta in forms of gaseous chemical elements (see Equation \ref{eq:metal-tracers}), 
while the rest are in form of dust grains (see Equation \ref{eq:dust-tracers}).
\item {\bf Toomre instability} in the circumnuclear disk.
We assume the stellar mass loss inherits the velocity of its host stars 
(thus it contributes as a momentum source term $\dot{\bf m}_{\rm S}$). 
Consequently,  a circumnuclear disk will form from infalling gas because of the angular momentum barrier
(e.g., \citealt{sarzi_sauron_2006,davis_atlas3d_2011,tadaki_gravitationally_2018}),
and inevitably become gravitationally unstable when gas accumulates 
continuously (i.e. the Toomre instability; \citealt{toomre_gravitational_1964}; 
\citealt{bertin_class_1999,goodman_self-gravity_2003,jiang_star_2011,frazer_gas_2019}).
Therefore, efficient mass accretion is possible by virtue of the Toomre instability, 
which will induce an inflowing mass flux ${\bf T}$ 
(calculated from Equations 13-15 in \citealt{gan_macer_2019}), 
and it is also capable in 
transferring angular momentum ($\Pi_{\rm T}$) outwards and dissipating energy ($\dot{E}_{\rm T}$),
and thus black hole accretion through the disk is possible.
\item {\bf Star formation.}
As another consequence of the gravitational instability, star formation is inevitable (see \S\ref{sec:jeans-criterion}).
Star formation is a sink for mass, momentum and energy
($\dot{\rho}_{\star}^{+}$, $\dot{\bf m}^{+}_{\star}$, $\dot{E}^{+}_{\star}$),
consuming a significant fraction of the accreting mass before it can reach
the central supermassive black hole.
We assume a top-heavy initial mass function (IMF) for the new stars,
as evidenced by the central young stellar populations in the Milky Way and M31, 
and as argued  by \citet{bartko_extremely_2010} 
(see also \citealt{levin_stellar_2003,lu_stellar_2013,calzetti_legacy_2015,palla_role_2019}).
Then, 
the distribution and age of the new stars sets the SNe II explosion rate.
\item {\bf Alpha viscosity} is included to mimic the angular momentum transfer due to the magnetorotational instability (MRI).
Besides the Toomre instability, the magnetorotational 
instability  can also transfer angular momentum ($\Pi_{\rm vis}$; 
\citealt{shakura_black_1973, balbus_instability_1998}).
Though it is inefficient on the length scale of interest,
the MRI will still take effect on the timescale we simulate.
\item {\bf AGN feedback} in forms of both radiation and wind.
We chose an inner boundary of 2.5 parsec, which is within the fiducial Bondi radius, so that 
we can track the mass inflow onto the galaxy center self-consistently.
We include a ``sub-grid" AGN model \citep{ostriker_momentum_2010,
yuan_active_2018} to allow for both the radiation and wind output 
from the black hole accretion disk based primarily on observations of AGN outbursts 
\citep{arav_quasar_2013,arav_evidence_2018,kara_glimpse_2018}. 
Finally, we  evaluate the AGN radiative feedback 
by calculating the metallicity-dependent radiative heating/cooling processes 
under the AGN irradiation 
($H - C$; $\nabla p_{\rm rad}$ is the radiation pressure; \citealt{sazonov_radiative_2005}), 
and evaluate the AGN mechanical feedback by injecting 
the wind output directly into the computational domain via the inner boundary.
The adopted AGN wind speed in the cold mode is set to be   
$6\times10^3$ km/s \citep{arav_evidence_2018} with an efficiency of $\epsilon_w = 1.5\times10^{-3}$.
The initial black hole mass is taken to be $10^{-3} M_{\rm \star}$ (cf. Table \ref{tab:galaxy-model}).
\item {\bf Cosmic accretion} through the galaxy outskirts. 
\textcolor{black}{
We allow CGM infall (cosmic accretion)} at the outer boundary 
by injecting time-dependent inflows based on cosmological simulations 
(\citealt{choi_physics_2017,brennan_momentum-driven_2018};
see \citealt{gan_macer_2019} for more details). 
The CGM infall and AGN wind feedback are implemented as boundary conditions; 
thus they won't appear in the equations above. 
The total CGM infall mass is now set to be $16\% M_\star$,
and the infall velocity is set at 50\% of the free-fall velocity.
\item {\bf Radiation pressure on dust grains.} Due to the high dust opacity, 
the radiation pressure on the dust grains (${\bf f}_{\rm rad,dust}$; 
see Equation \ref{eq:rad-pressure-on-dust}) can be much higher than that on the electrons. 
In our simulations, we assume the dust grains are coupled with gas dynamics, 
and so the radiation pressure on dust will be important mechanism 
to regulate the gas flow, especially during the outbursts of the AGN.
\item {\bf Dust cooling of hot gas}. In the densest zone in the circumnuclear disk 
(where the gas density reaches $>10^5$ atom/cm$^3$), most of the metals are in dust grains. 
There the frequent collisions between gas and dust particles ($H^{\rm gas}_{\rm dust,collision}$; 
see Equation \ref{eq:dust-collisional-heating}) will cool down the gas efficiently on a timescale 
as short as $\sim1$ year. Therefore, dust cooling is an important mechanism in the simulation 
to allow the gas to be further cooled down below $10^4$ K.
\end{enumerate}

The ISM hydrodynamical equations, together with the tracers for metals and dust, are solved by 
using the grid-based
\texttt{Athena++} code \citep[version 1.0.0;][]{stone_athena:_2008} in spherical coordinates. 
We assume axi-symmetry but allow rotation. 
The outer boundary is chosen as 250 kilo-parsec to enclose the whole massive elliptical galaxy, 
the inner boundary $R_{\rm in}$ is set to be 2.5 parsec to resolve the Bondi radius.
We use a logarithmic grid ($\Delta r_{\rm i+1}/ \Delta r_{\rm i} = 1.1$) to 
divide the radial axis into 120 discrete cells. The azimuthal angle $\theta$ 
is divided into 30 uniform cells and covers an azimuthal range from $0.05\pi$ to $0.95\pi$. 
The numerical solver for the gas dynamics is composed 
by the combination of the HLLE Riemann Solver, the PLM reconstruction 
and the second-order van Leer integrator. 
We start the simulations when the initial stellar population is 2 Gyr old, 
\textcolor{black}{i.e., after the major phase of galaxy formation, 
we therefore maintain the galaxy structure and internal kinematics fixed during the simulations.}

\section{Chemical Abundances} \label{sec:metals}
In \citet{gan_adding_2019}, we track the chemical evolution of \textcolor{black}{H, He and} metals 
by using 12 tracers $X_{\rm i}$ (i=1, 2, ..., 12; \textcolor{black}{mass of each element per unit volume}) 
for H, He, C, N, O, Ne, Mg, Si, S, Ca, Fe and Ni, respectively 
\textcolor{black}{(\textcolor{black}{inspired by}  \citealt{choi_physics_2017}, \citealt{eisenreich_active_2017}
for large scale simulations using SPH).}
We solve 12 additional continuity equations of the tracers, assuming the chemical species 
co-move once after they are injected into the ISM, i.e.,
\begin{equation} \label{eq:metal-tracers}
   \frac{\partial X_{\rm i}}{\partial t} + \nabla\cdot(X_{\rm i}{\bf v}) + \nabla\cdot{\bf T}_{\rm i} 
        = \dot{X}_{\rm \star,i} + \dot{X}_{\rm I,i} + \dot{X}_{\rm II,i} - \dot{X}_{\rm \star,i}^{+},
\end{equation}
where
\begin{equation} 
   {\bf T}_{\rm i} = (X_{\rm i}/\rho_{\rm gas}) \cdot {\bf T}, \quad
   \dot{X}_{\rm \star,i}^{+} = (X_{\rm i}/\rho_{\rm gas}) \cdot \dot{\rho}^{+}_{\star},
\end{equation}
and ${\bf v}$ is obtained by solving the hydrodynamical equations as usual. 
Here ${\bf T}_{\rm i}$ is the inflowing mass flux of chemical species i induced by the Toomre instability.
The products of stellar evolution in Equation 4, i.e., AGB stars ($\dot{X}_{\rm \star,i}$), 
SNe Ia ($\dot{X}_{\rm I,i}$) and SNe II ($\dot{X}_{\rm II,i}$),  \textcolor{black}{serve as sources} of metals 
with each having different compositions (see Table \ref{tab:chemical-composition}).  
The time-dependent metal yields above are calculated assuming standard stellar physics \citep{saitoh_chemical_2017,gan_adding_2019}.
The advection terms $\nabla\cdot(X_{\rm i}{\bf v})$ and $\nabla\cdot{\bf T}_{\rm i}$ 
\textcolor{black}{describe the transport and the mixing of} ISM with different metal abundances.
Star formation ($\dot{X}_{\rm \star,i}^{+}$) is treated as a sink of metals, 
but does not to change the \textcolor{black}{local} relative abundances.

On the galaxy outskirts, we include the cosmic accretion (i.e., CGM infall).
The CGM is low-metallicity by construction (see Table \ref{tab:chemical-composition}), 
therefore it will dilute the metallicity as it falls into the galaxy and mixes with the ISM there.  
At the inner boundary of the simulation domain, we also recycle the metals via the BAL winds which are injected back 
to the galaxy by the central AGN.
The metal abundance of the BAL winds is determined by the chemical composition of the inflow at our innermost grid point.
\citep{ciotti_cooling_1997, gan_macer_2019}. 
Finally, we assume the abundance of the initial ISM (cf. \S\ref{sec:galaxy-model}) is of $Z \sim 1.54 Z_\odot$ 
(where $Z_\odot$ is the solar metallicity;  \citealt{asplund_chemical_2009}).
The initial abundance \textcolor{black}{of $1.54 Z_\odot$}  
is adopted from the zoom-in cosmological simulations by \citet{choi_physics_2017}
for elliptical galaxies of similar mass as in our galaxy model \textcolor{black}{(see also \citealt{eisenreich_active_2017})}.

It is a goal of this paper to evaluate the depletion of metals onto dust, especially in the cold gas component; 
the fractional depletion of refractory elements onto dust grains may be a large correction.  
The total infrared radiation from the dust may be comparable to or greater than the X-rays from the hot gas.

\begin{table}[ht] 
\caption{Mass fraction of the elements from various sources}
\label{tab:chemical-composition}
\begin{center}
\footnotesize
\begin{tabular}{lcccccc}
\hline\hline
{  } & {AGBs$^a$} &{SNe Ia$^b$} & {SNe II$^c$} & {CGM$^d$} & {Solar$^e$} & {Dust$^f$}\\
\hline
H & 0.712859 & 0.000000 & 0.517998 & 0.746829 & 0.735972 & 0.000000 \\
He& 0.267027 & 0.000000 & 0.334833 & 0.251177 & 0.250628 & 0.000000 \\
C & 0.002939 & 0.002247 & 0.010310 & 0.000362 & 0.002411 & 0.171014 \\
N & 0.001834 & 0.000002 & 0.003618 & 0.000106 & 0.000706 & 0.000000 \\
O & 0.008723 & 0.074648 & 0.081750 & 0.000871 & 0.005849 & 0.293720 \\
Ne& 0.001895 & 0.002639 & 0.025177 & 0.000191 & 0.001271 & 0.000000 \\
Mg& 0.001064 & 0.011234 & 0.007615 & 0.000107 & 0.000713 & 0.117874 \\
Si& 0.000470 & 0.212119 & 0.003912 & 0.000101 & 0.000676 & 0.115942 \\
S & 0.001010 & 0.084995 & 0.008612 & 0.000047 & 0.000315 & 0.042512 \\
Ca& 0.000097 & 0.010865 & 0.000497 & 0.000010 & 0.000065 & 0.012560 \\
Fe& 0.001973 & 0.546927 & 0.005402 & 0.000198 & 0.001322 & 0.234783 \\
Ni& 0.000107 & 0.054323 & 0.000275 & 0.000001 & 0.000072 & 0.011594 \\
\hline
 Z$^g$& 0.020114 & 1.000000 & 0.147168 & 0.001994 & 0.013400 & 1.000000 \\
\hline \hline
\end{tabular}
\end{center}

\footnotesize
\hangindent 0.75em
$^a$ averaged metal abundance of AGB winds over the time span from $t_{\rm age} = 2$
         to \textcolor{black}{13.7} Gyr, i.e., $<{\dot{X}_{\rm \star, i}}/\dot{\rho}_{\rm \star}>$ 
         \citep{karakas_updated_2010}; 
         
$^b$ metal abundance of SNe Ia ejecta, i.e., {$\dot{X}_{\rm I, i}/\dot{\rho}_{\rm I}$} 
	\citep{seitenzahl_three-dimensional_2013};
	
$^c$ metal abundance of SNe II ejecta, i.e., {$\dot{X}_{\rm II, i}/\dot{\rho}_{\rm II}$} 
	\citep{nomoto_nucleosynthesis_2013};
	
\hangindent 0.75em	
$^d$ metal abundance of the low-metallicity infalling CGM  which is made of 1/4 of primordial gas
        and 3/4 low metallicity gas of 0.2 solar abundance; 
               
$^e$ solar abundance \citep{asplund_chemical_2009};

$^f$ metal composition of dust grains, i.e., $<\tilde{D}_{\rm i}/\tilde{\rho}_{\rm dust}>$;

$^g$ metallicity Z, i.e, mass fraction of all chemical species except H and He. 
\end{table}

\section{Grain physics} \label{sec:grain-physics}
\subsection{Grain Hydrodynamics}
Dust grains are injected into the ISM by both AGB winds and
supernovae. Grains are quickly coupled to gas via collisions, and so
are passively advected. Collisions with hot gas atoms can erode
grains, while dust in cold gas can grow via accretion of metals. In
parallel to the treatment of metals (Equation~
\ref{eq:metal-tracers}), we implement these processes in a dust
continuity equation \citep[see also][]{hensley_grain_2014}:

\begin{equation} \label{eq:dust-tracers}
 \begin{aligned}
   \frac{\partial \rho_{\rm dust}}{\partial t} +  \nabla\cdot( \rho_{\rm dust}{\bf v}) &+ \nabla\cdot{\bf T}_{\rm dust} 
        = \dot{\rho}_{\rm \star,dust} \\&- \dot{\rho}_{\rm \star,dust}^{+} + \dot{\rho}_{\rm dust, gg} - \dot{\rho}_{\rm dust, gd} ,
 \end{aligned}
\end{equation}
where
\begin{equation} 
   {\bf T}_{\rm dust} = (\rho_{\rm dust}/\rho_{\rm gas}) \cdot {\bf T}, \quad
   \dot{\rho}_{\rm \star,dust}^{+} = (\rho_{\rm dust}/\rho_{\rm gas}) \cdot \dot{\rho}^{+}_{\star},
\end{equation}
are the mass advection and sink terms due to the Toomre instability and star formation, respectively. 

The mass density $X_{\rm i}$ of each element i is given by
Equation~\ref{eq:metal-tracers}. We partition this into a gas phase
density $G_{\rm i}$ and a density in dust grains $D_{\rm i}$, i.e.,

\begin{equation} \label{eq:mass-consrv-metal}
  X_{\rm i} =  G_{\rm i} +  D_{\rm i}
  ~~~.
\end{equation}
$X_{\rm i}$ and $\rho_{\rm dust}$ can be solved from the continuity equations 
\ref{eq:metal-tracers} and \ref{eq:dust-tracers}, respectively. To
derive $G_{\rm i}$ and $D_{\rm i}$, we assume for simplicity that
all grains have a fixed composition. Table~\ref{tab:chemical-composition} lists the
assumed mass fraction of each metal in dust $<\tilde{D}_{\rm
  i}/\tilde{\rho}_{\rm dust}>$. Thus, $D_{\rm i}$ is given by

\begin{equation} \label{eq:dust-metal-composition}
D_{\rm i} \equiv \rho_{\rm dust} \cdot <\tilde{D}_{\rm
  i}/\tilde{\rho}_{\rm dust}>
~~~,
\end{equation}
with $G_{\rm i}$ following from
Equation~\ref{eq:mass-consrv-metal}. This formulation allows us to
track the time evolution  and the spatial distribution of each
chemical species in the gaseous ISM and in dust.

The source term $\dot{\rho}_{\rm \star, dust}$ in
Equation~\ref{eq:dust-tracers} describes the injection of dust grains
into the ISM from dying stars, including both AGB winds and
\textcolor{black}{supernovae}. Specifically,
\begin{equation} 
        \dot{\rho}_{\rm \star, dust} =  \dot{\rho}_{\rm \star}\cdot R_{\rm \star, dust}
                                                 +  \dot{\rho}_{\rm I}\cdot R_{\rm I,dust} 
                                                 + \dot{\rho}_{\rm II}\cdot R_{\rm II,dust},
\end{equation}
where $R_{\rm \star, dust}$, $R_{\rm I, dust}$, and $R_{\rm II, dust}$ are the maximum ratios 
of dust mass to the total ejecta of AGB stars, SN Ia, and SN II, respectively, 
when the total available metals are transformed to the dust phase, 
e.g.,
\begin{equation} \label{eq:stellar-ejecta-dust-ratio}
R_{\rm \star, dust} =  1 - R_{\rm \star, gas}
=\min_{\textcolor{black}{i=1...12}}\left(\frac{\dot{X}_{\rm \star,
      i}/\dot{\rho}_{\rm \star}}{D_{\rm i}/\rho_{\rm dust}}\right)
~~~.
\end{equation}
\textcolor{black}{That is, we assume that each of these sources injects as much
dust of our prescribed composition --- 
numerically, we ``make'' dust grains 
from the stellar metal yields
until any of the metal ingredients runs out,  then no more dust grains could be made 
according to the fixed dust chemical composition, i.e., 
each planetary nebula makes as much dust as its ejected metals permit.}
We calculate
($R_{\rm I, dust}$, $R_{\rm I,  gas}$) and ($R_{\rm II, dust}$,
$R_{\rm II, gas}$) in a similar way. 

Finally, we track the mass exchange between the gaseous ISM and the dust 
by allowing grain growth $\dot{\rho}_{\rm dust, gg}$ and grain
destruction $\dot{\rho}_{\rm dust, gd}$. These are described in more
detail in the following sections.

\subsection{Grain Destruction}
In hot gas, impacting ions can erode dust grains on short
timescales \citep{Draine_physics_1979}. Thermal sputtering is
expected to be the dominant grain
destruction process in the hot ISM of an elliptical galaxy. Following
\citet{hensley_grain_2014}, we adopt an analytic approximation for the
sputtering rate that depends only on the gas density and temperature
\citep{draine_physics_2011}. Letting $a$ be the grain radius, $T_6$
the gas temperature in units of $10^6$ K, and $n_{\rm H}$ the proton
number density in units of $\mathrm{cm}^{-3}$,

\begin{equation}
\ \dot{a} = -\frac{10^{-6} n_{\rm H}}{1 +
  T_6^{-3}}\ \mu\mathrm{m}\
\mathrm{yr}^{-1}
~~~.
\end{equation}
We assume that the grain size distribution is everywhere given by the
Mathis-Rumpl-Nordsieck \citep[MRN;][]{mathis_size_1977} size distribution, i.e.,

\begin{equation}
\label{eq:mrn_dist}
\ \frac{{d}n_d}{{d}a} = \frac{A a_{\rm max}^{2.5}}{a^{3.5}}
\end{equation}
where $(dn/da)da$ is the number density of grains with size between
$a$ and $a+da$. $A$ is a normalization factor that can be
determined from the total dust mass density $\rho_{\rm dust}$ via
\begin{equation}
\label{eq:rho_gd}
\rho_{\rm dust} = \frac{8 \pi}{3} A \left(1 - \sqrt{a_{\rm min}/a_{\rm max}} \right) 
   \rho_{\mathrm{grain}}a_{\mathrm{max}}^3,
\end{equation}
where $\rho_{\mathrm{grain}}$ is the mass density of a dust grain,
taken to be 3.5\,g\,cm$^{-3}$ as typical for a silicate grains \citep{draine_optical_1984}.
Guided by \citet{mathis_size_1977} and \citet{weingartner_dust_2001}, we
adopt $a_{\rm min} = 0.005 \mu$m and $a_{\rm max} = 0.3 \mu$m. The
total destruction rate  is obtained by integrating the mass
destruction rate of grains of radius $a$ over the size distribution, i.e.,
\begin{equation}
\label{eq:rhodot_gd}
\ \dot{\rho}_{\rm dust,gd} = 8 \pi A \rho_{\rm grain}
\left(\frac{a_{\rm max}}{a_{\rm min}}\right)^{0.5}a_{\rm max}^2\dot{a}.
\end{equation}

This implementation of thermal sputtering \textcolor{black}{implicitly} assumes that the
all of the gas in a cell can be described by a single
temperature. However, cool, dense gas can be shock heated by
supernovae, creating conditions favorable for non-thermal sputtering
and grain-grain collisions at spatial scales not resolved by the
simulation \citep{seab_shock_1983}. Grain destruction by
supernova-driven shocks has been implemented in other studies by
coupling the grain destruction timescale to the supernova rate
\citep[e.g.,][]{mckinnon_dust_2016,zhukovska_modeling_2016}. Given the
nature of the hot ISM of an elliptical galaxy and the necessarily
phenomenological approach needed to incorporate this subgrid physics,
we do not consider this destruction mechanism in this work. However,
as we discuss in Section~\ref{sec:IR-obs-low-redshift}, this may lead to
an overestimate of the dust content in cool gas in the simulation. 

\subsection{Grain Growth}
When the gas density is sufficiently high, dust grains can grow from
accretion of metal atoms from the gas phase (\citealt{hensley_grain_2014}, and the references therein). 
Assuming a sticking efficiency $f$ and average speed $v_Z$ of these metal atoms, the
the grain growth rate due to collisions is $f \rho_Z v_Z 4 \pi a^2 n_d$,
where $\rho_Z$ is the mass density of the metals available for grain growth, i.e.,
\begin{equation} 
\rho_{\rm Z} = {\rho}_{\rm gas}\cdot
\min_{\textcolor{black}{i=1...12}}\left(\frac{{G}_{\rm i}/{\rho}_{\rm
      gas}}{D_{\rm i}/\rho_{\rm dust}}\right).
\end{equation}
In this formualtion, $\rho_{\rm Z} \to 0$ when any of the atomic
constituents of dust (see Table~\ref{tab:chemical-composition}) has been
significantly depleted from the gas phase, thereby limiting the
efficiency of grain growth.

Integrating over the grain size distribution, we obtain
\begin{equation}
\ \dot{\rho}_{\rm dust,gg} = f \rho_Z \frac{c_s}{\sqrt{\mu_Z}} 8 \pi  A a_{\rm max}^2
\left(\frac{a_{max}}{a_{\rm min}}\right)^{0.5}.
\end{equation}
In the equation above,  we have made the approximation $v_Z = c_s/\sqrt{\mu_Z}$, 
where $c_s$ is the sound speed of the gas and $\mu_{Z}$ is the mean atomic mass of
the metal atoms. For simplicity, we set constant $f=0.2$ and $\mu_{Z}=16$.

\subsection{Dust Cooling of Hot Gas}
It is known that dense ISM will be cooled down efficiently via inelastic collisions 
with dust grains \citep{ostriker_dust_1973,smith_time-dependent_1996,draine_physics_2011}. 
The cooling rate of gas per unit volume per unit time can be written as, 
\begin{equation}  \label{eq:dust-collisional-heating}
 \begin{aligned}
      H^{\rm gas}_{\rm dust,collision}  = &\int_{a_{\mathrm{min}}}^{a_{\mathrm{max}}} \! {d}a \frac{{d}n_d}{{d}a} 
                                                                \cdot 4 \pi a^2  \cdot \frac{1}{2}\rho_{\rm gas} c_s^3 \\
                                                             &\quad \cdot \left[1+ \frac{T_4}{1+T_4}\left({\frac{m_p}{m_e}}\right)^{\frac{1}{4}} \right],
 \end{aligned}                
\end{equation}
 where $T_4$ is the gas temperature in units of $10^4$ K.
 \textcolor{black}{The factor in the brackets is a correction for the energy exchange considering the charge of dust grains 
and that electrons dominate the collisions when the gas temperature is high. }

In the circumnuclear disk, both the gas density and dust density can be very high, 
which makes the process above very efficient --- its timescale could be as short as $\lesssim1$ year. 
Therefore, it is a stiff sink term in the energy equation of the gas dynamics, 
and it is numerically expensive to resolve the timescale explicitly.  
To overcome this difficulty, we assume thermal equilibrium 
when the timescale is shorter than the timestep given by the  Courant$–$Friedrichs$–$Lewy (CFL) condition --- 
we calculate the equilibrium temperature while assuming the thermal pressure is fixed. 
After that, we evaluate the energy loss according to the difference 
between the instantaneous gas temperature and the equilibrium temperature.

\subsection{Radiation Pressure on Dust Grains}
It is known that 
absorption by dust grains is the dominant sources of AGN obscuration, 
often exceeding the opacity due to electron scattering by orders of magnitude.
The dust opacity in the UV,  optical and infrared bands can be written roughly, 
\begin{equation} \label{eq:dust-opacity}
 \begin{aligned}
  &\kappa_\mathrm{dust,opt} = 300\times
            \left(\frac{\rho_{\rm dust}}{\rho_{\rm gas}}\right)
            \left(\frac{\rho_{\rm dust}}{\rho_{\rm gas}}\right)^{-1}_{\rm MW} 
            ~{\rm cm}^2/{\rm g}, \\
  &\kappa_\mathrm{dust,UV} = 4 \kappa_\mathrm{dust,opt}, \quad
  \kappa_\mathrm{dust,IR} \ \ =\kappa_\mathrm{dust,opt}/{150}.
 \end{aligned}
\end{equation}
where we have scaled the opacities according to the dust-to-gas ratio normalized by the MW value.
For simplicity, we only consider the absorption of UV photons from the central AGN by dust grains. 
The photon energy absorbed by dust grains per unit volume per unit time can be written as 
\begin{equation}  \label{eq:dust-radiative-heating}
      H^{\rm AGN,UV}_{\rm dust,abs}  = \rho_{\rm dust} \kappa_{\rm dust, UV} \cdot \frac{L_{\rm AGN, UV}}{4 \pi r^2} \exp(-\tau_{\rm dust, UV}) 
\end{equation}

\begin{table}[ht]
\caption{Parameters of the initial galaxy model}\label{tab:galaxy-model}
\centering
\footnotesize
\begin{tabular}{llll}
\hline\hline
{Galaxy property} & {Value}  & {Description}\\
\hline
$L_B$               &  $1.25\times10^{11} L_{B\odot}$        &   total stellar luminosity              \\ 
$M_\star$          &  $6.1\times10^{11}M_\odot$              &   total stellar mass\\
$R_e$               &  $11.36$ kpc                                       &  projected half light radius\\
$M_g$               &  $1.22\times10^{13}M_\odot$            & total mass of the galaxy\\
$\sigma_\star$  & $260$ km/s                                       & 1D velocity dispersion at 0.1$R_e$ \\
$\epsilon$         & 0.37                                                   &  ellipticity\\
$M_{\rm BH}$   &  $6.1\times10^8 M_\odot$                 &  initial mass of central black hole   \\
\hline \hline 
\end{tabular}
\end{table}

\noindent
where  $L_{\rm AGN, UV}$ is the AGN luminosity in the UV band. $\tau_{\rm dust, UV}$ 
is the optical depth of dust in the UV band along radial directions, i.e., 
\begin{equation}  \label{eq:dust-opt-depth-uv}
      \tau_{\rm dust, UV} =  \int_{r_{\rm in}}^{r} \rho_{\rm dust} \kappa_{\rm dust, UV} \cdot {d} r^\prime.
\end{equation}
The calculation of the optical depth above is very computationally expensive 
because it could not be parallelized. For the sake of computational efficiency, 
we update the optical depth only every 100 timesteps.

The dust grains gain momentum when they absorb the photons, 
which results in the radiation pressure on dust grains. Compared to electron scattering, 
the radiation pressure due to irradiation by AGN could be very significant especially during bursts. 
Since we assume the dust grains are coupled with gas dynamics, 
there will be a net acceleration to the gas as follows \citep{draine_physics_2011},
\begin{equation}  \label{eq:rad-pressure-on-dust}
        {\bf  f}_{\rm rad,dust}   = \frac{\rho_{\rm dust} \kappa_{\rm dust, UV}} {\rho_{\rm gas}+\rho_{\rm dust}}  \cdot 
                                             \frac{L_{\rm AGN, UV}}{4 \pi r^2 c} \exp(-\tau_{\rm dust, UV}) \cdot \hat{\bf r}.
\end{equation}

\subsection{\textcolor{black}{Initial and Boundary Conditions for Dust}}
At the inner boundary, we keep track of the dust accreted onto the black hole accretion disk. 
Since the temperature of the black hole accretion disk (even in the cold mode) is high enough 
\textcolor{black}{so} that no dust can survive,  
we return the metals in dust grains (which are accreted onto the center) 
back to the gas phase where they are ejected in the BAL winds.
For simplicity, we assume the BAL wind is dust-free, as it can be easily heated up by shocks 
which will make it difficult for dust to survive. 

At the outer boundary,  we allow for inflows of low-metallicity material ($Z=0.15Z_\odot$; cf. Table \ref{tab:chemical-composition}) to mimic the CGM infall.  
For simplicity, we assume that the CGM is dust-free.

Finally, we assume that the initial ISM is dust-free as its temperature is high.

\section{Adjustments to the Numerical Models} \label{sec:model-updates}

\subsection{Initial Galaxy Model} \label{sec:galaxy-model}

The simulation in this paper is designed for
an isolated, massive elliptical galaxy, such as NGC 5129.
The galaxy model is built using the procedure described in \citet{gan_macer_2019},
and its main properties are summarized in Table \ref{tab:galaxy-model} 
(see also Pellegrini et al. 2021, in preparation; \citealt{bentz_black_2018});
The only difference is a radius-dependent description of the Satoh's k-parameter 
that is now parametrized as
\begin{equation} \label{eq:rotation-profile}
     k(r) = k_0 + (k_{est}-k_0)\cdot {{r\over r_0+r}},
\end{equation}
where $k_0=0.42, k_{est}=0.05$, {$r_0=40$ kpc}. 
These parameters produce a rather flat
rotation curve, that approximates $v_{\rm rot}\sim 100$ km s$^{-1}$ when $r\lesssim R_e$,
and then decreases outward 
\textcolor{black}{in order to reproduce qualitatively the main features of the rotation profile of NGC 5129}.

In our standard simulations we start with a gas free galaxy allowing infall and stellar evolution to provide all of the subsequent ISM. 
In our results section (\S\ref{sec:resolution-study}) we will also describe the results of a perhaps more plausible initial state where the gas/star mass ratio at the beginning of the simulation is $\sim30\%$.

\subsection{Star Formation Criterion} \label{sec:jeans-criterion}
In our model, we have two physical channels for star formation, 
i.e., one via the Toomre instability, when the disk is gravitationally unstable 
\citep{toomre_gravitational_1964,goodman_self-gravity_2003,jiang_star_2011},
and the other via radiative cooling when it is Jeans unstable. 
To ensure it is gravitationally unstable at the site of star formation, 
we 
allow star formation due to radiative cooling 
only when the local Jeans radius is smaller than the cell height. The Jeans radius reads,
\begin{equation} \label{eq:jeans-radius}
   R_{\rm Jeans} =3 \times \left(\frac{c_s}{2~{\rm km/s}}\right)  \sqrt{\frac{10^3~{\rm cm}^{-3}}{n_{\rm H}}} \quad {\rm parsec}.
\end{equation}

We have not made allowance for shear or turbulence velocities {\it within} cells; 
these effects would reduce star formation in the innermost regions.

\subsection{Cooling Radius of Supernovae}
In the densest zone on the circumnuclear disk, the cooling radius of SN II feedback can be very small, 
i.e., the SN II heating will fade before it can affect its distant surroundings. 
Therefore, we allow SN II heating only when its cooling radius is larger than the cell height. 
The cooling radius of SN II is approximately 
\begin{equation} \label{eq:SNII-cooling-radius}
   R_{\rm cool} = 5 \times E_{51}^{1/5} n_H^{1/5} \left(\frac{t_{\rm cool}}{10^3~{\rm year}}\right)^{2/5} \quad {\rm parsec},
\end{equation}
where $E_{51}$ is the total energy output of a SN II explosion in  units of $10^{51}$ erg. 
$t_{\rm cool} = {\rm min}(t_{\rm low}, t_{\rm high})$ is the time of expansion pf supernova remnant before cooling. 
$t_{\rm low}$ and $t_{\rm high}$ are the estimates from the low and high temperature branches, respectively 
(see \citealt{kim_momentum_2015}),
\begin{equation} 
 \begin{aligned} 
   &t_{\rm low}  = 4\times10^4 ~E_{51}^{0.22} n_H^{-0.55} \quad {\rm yr}, \quad\quad \\
   &t_{\rm high} = 4\times10^5 ~E_{51}^{0.38} n_H^{-0.55} \quad {\rm yr}.
 \end{aligned}
\end{equation}

\subsection{Runaway Stars}
In our previous model \citep[e.g.,][]{gan_adding_2019}, 
we assumed a top-heavy IMF for the star formation in the circumnuclear disk
\citep{goodman_supermassive_2004,bartko_extremely_2010,lu_stellar_2013,palla_role_2019}, 
i.e., $\sim60\%$ of the mass in new stars is in massive stars, of which a large fraction are in binary systems. 
Thus, there is a significant \textcolor{black}{probability} for the massive stars to be kicked out of the circumnuclear disk, 
typically, with a runaway speed greater than 30 km/s. 
Finally, the runaway stars explode outside the circumnuclear disk and heat the ISM there.  
Therefore, the runaway stars may be an additional heating source for the ISM in the central galactic regions,
which is a direct consequence of the star formation in the circumnuclear disk. 
Thus, following \citet{eldridge_runaway_2011} (their Figure 2), 
we assume 40\% of the massive stars are runaway stars, 
and sample the escaping speed of the runaway stars 
according to the fitting formula below for the observed cumulative distribution function (CDF),
\begin{equation}  \label{eq:runaway-star-speed}
   v_{\rm runaway} = 30 - 100\log(1-{\rm CDF})     \quad\quad   {\rm km/s},
\end{equation}
i.e., no runaway stars have $v_{\rm runaway}<30$ km/s, 50\% of the stars have that $v_{\rm runaway}<60$ km/s, 
90\% of the stars have $v_{\rm runaway}<130$ km/s, etc. 
We set a fixed travel time of the runaway stars to $5\times10^6$ year, 
while the direction of the runaway stars is sampled randomly with equal probability over the solid angle of $4\pi$.

\subsection{Mass-Dependent Explosion Energy of SNe II}
Previously, we assumed that all SNe II inject a single characteristic explosion energy of $10^{51}$ erg/s. 
However, the explosion energy of a type II supernova should be dependent 
on the mass of its progenitor. In this paper, we adopt the mass-dependent SN II explosion energy 
in \citeauthor{morozova_measuring_2018} (\citeyear{morozova_measuring_2018}; their Figure 6), 
 fitting it with the formula below, 
\begin{equation}  \label{eq:snii-energy}
   \log(E_{\rm SNII}) = 50.6 + 2.2 \log(M/10M_\odot)    \quad\quad   {\rm erg/s}.
\end{equation}

\subsection{Kinetic Feedback of SNe II}
We now allow SN II kinetic feedback, i.e., we keep the vertical momentum component 
of the SN II ejecta, which corresponds to 1/16 of the total SN II energy, 
while the other 15/16 of the energy is still in form of thermal energy.

\begin{figure*}[htb]
\centering
\includegraphics[width=0.625\textwidth]{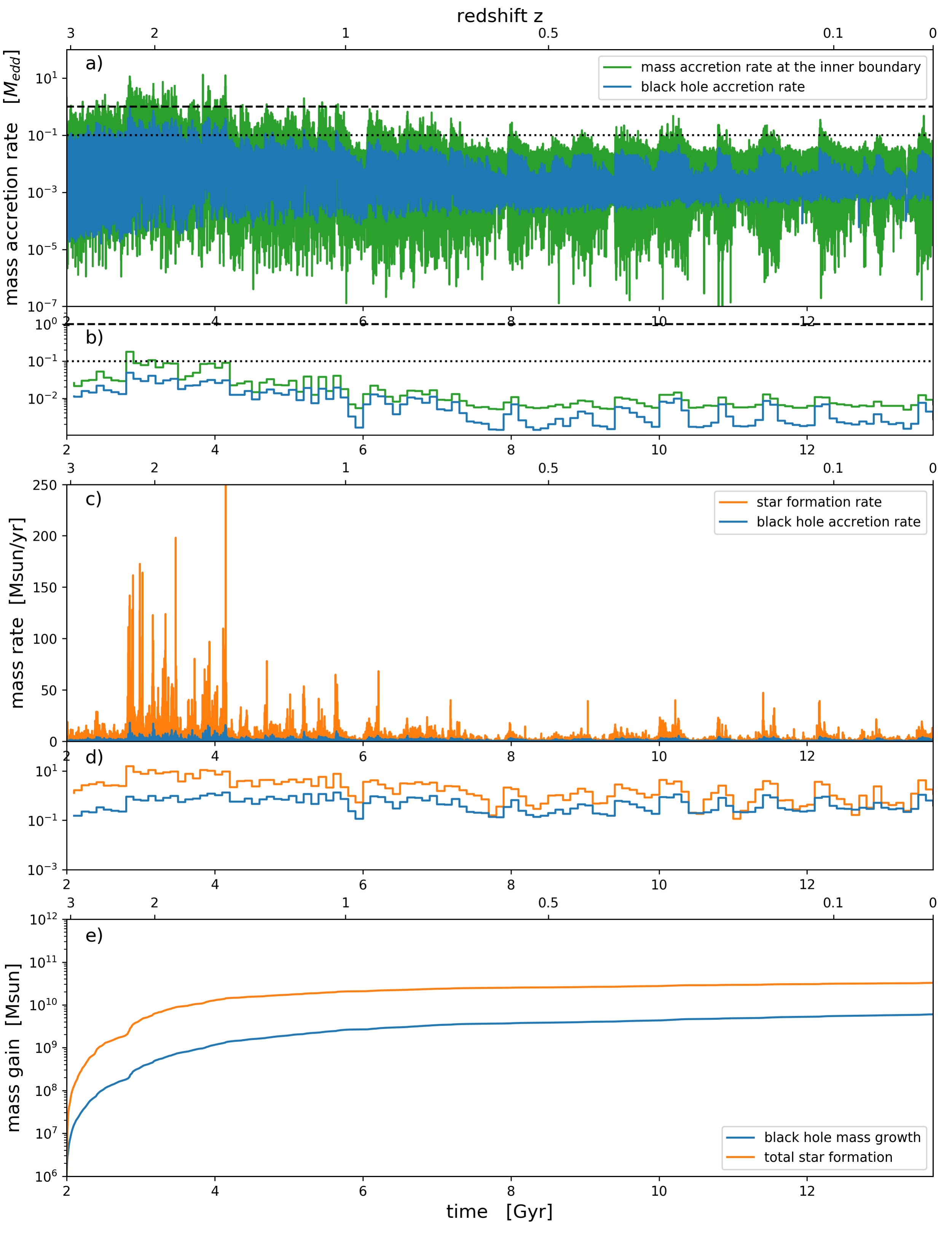}
\caption{History of the mass supply onto the galaxy center, AGN activities and star formation in the simulation. 
\textcolor{black}{The time resolution in Panels (a, c, e) is $\sim10~{\rm yr}$, while Panels (b, d) average over a $10^8~{\rm yr}$ timescale.}}
\label{fig:sim-overview}
\end{figure*}
\begin{figure*}[htb]
\centering
\includegraphics[width=0.45\textwidth]{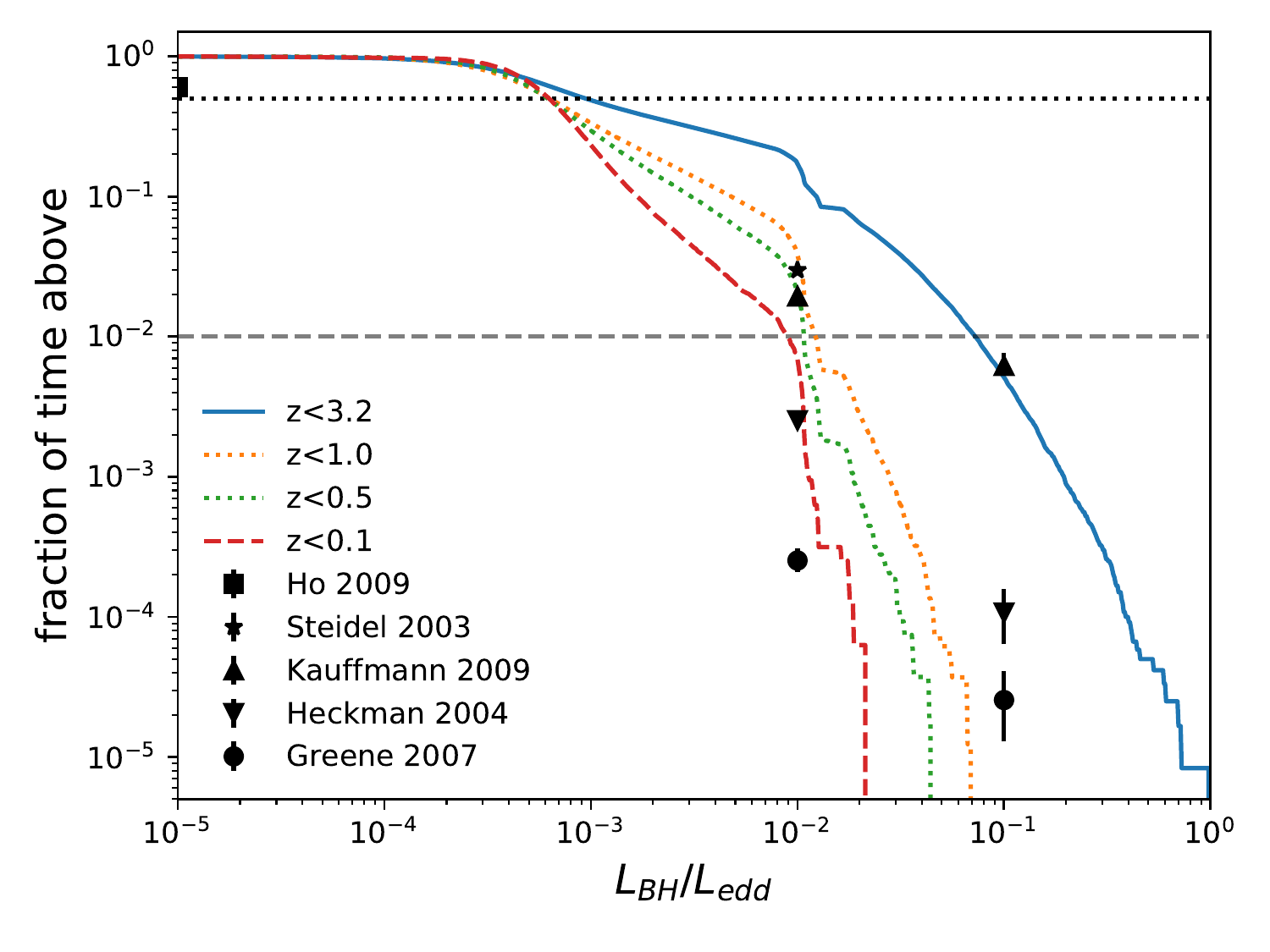}
\includegraphics[width=0.45\textwidth]{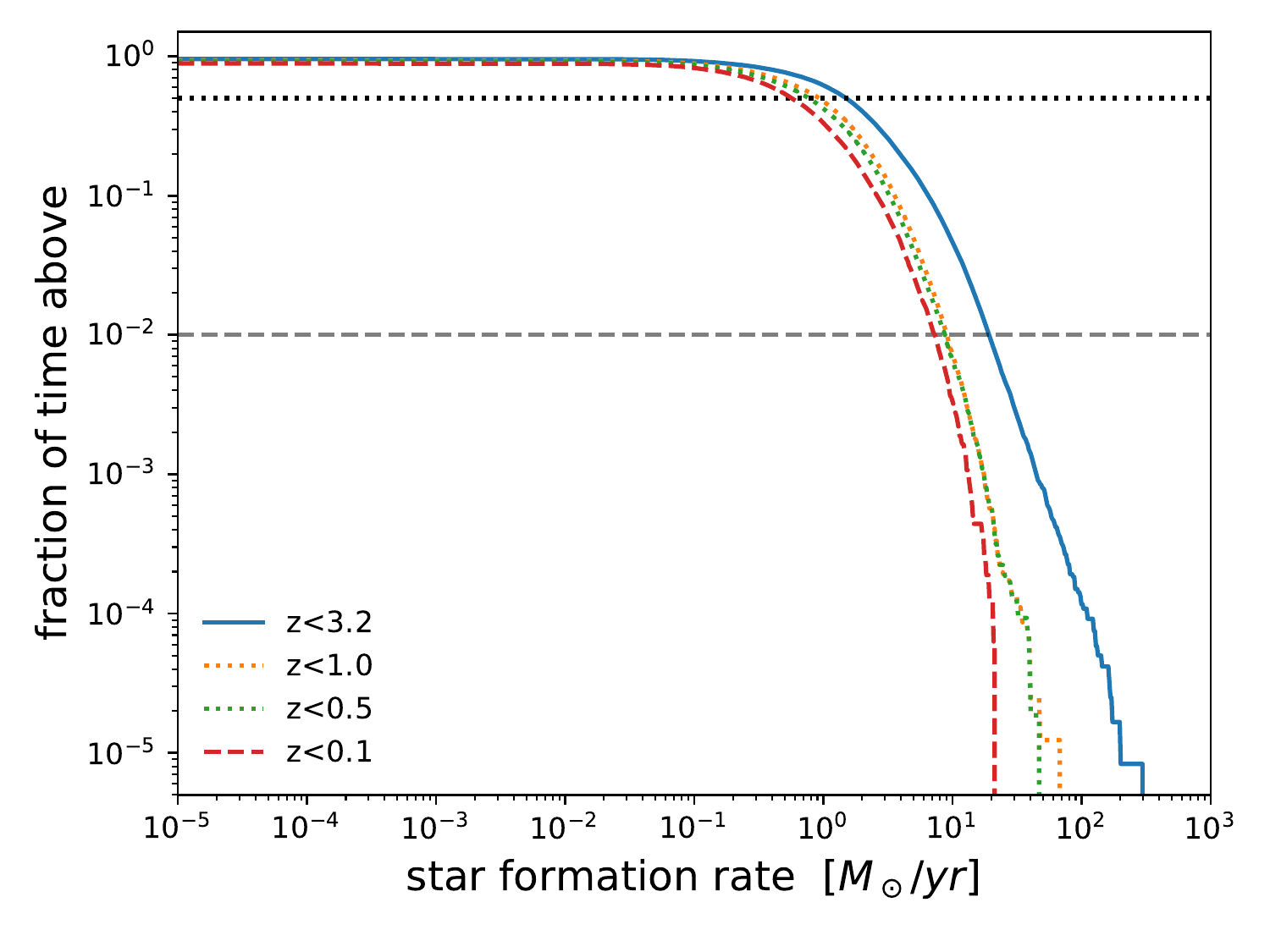}
\caption{Duty cycle of the AGN activities (left panel) and star formation (right panel), 
i.e., the fraction of time when the quantities are above given values. 
The horizontal dotted lines are at $50\%$ and dashed at 1\%. 
The lines in colors are the simulation results below given red shift.
The points in the left panel are observational constraints: 
The star is a constraint compiled from high-redshift observations by \citet{Steidel_lyman_2003}; 
The squares, circles, and upward- and downward- pointing triangles are from \citet{Ho_origin_2009}, 
\citet{Greene_mass_2007}, \citet{Kauffmann_feast_2009} and \citet{Heckman_presentday_2004}, respectively, 
which are all compiled from low-redshift observations. \\ \\
}
\label{fig:duty-cycle}
\includegraphics[width=0.3\textwidth]{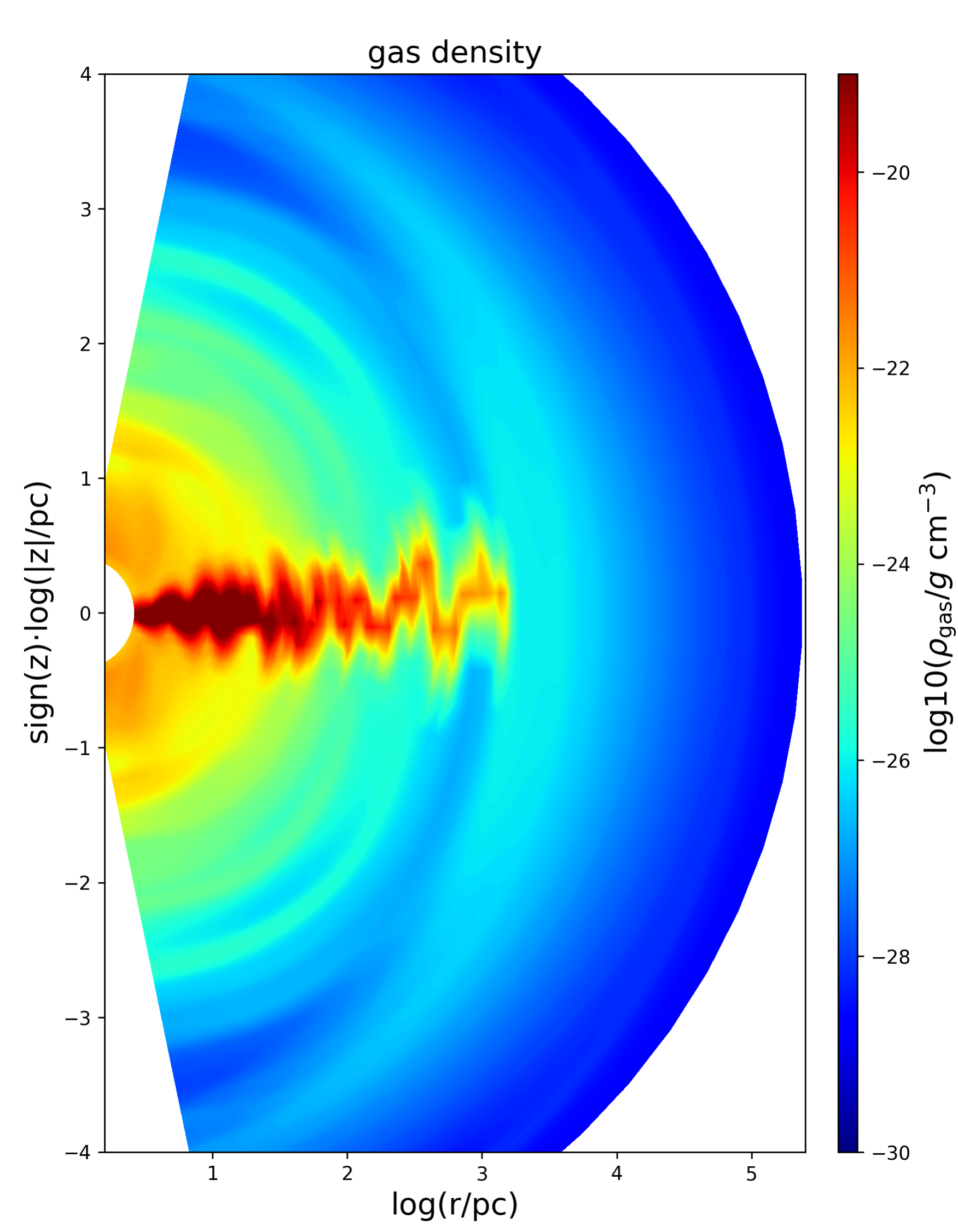}
\includegraphics[width=0.3\textwidth]{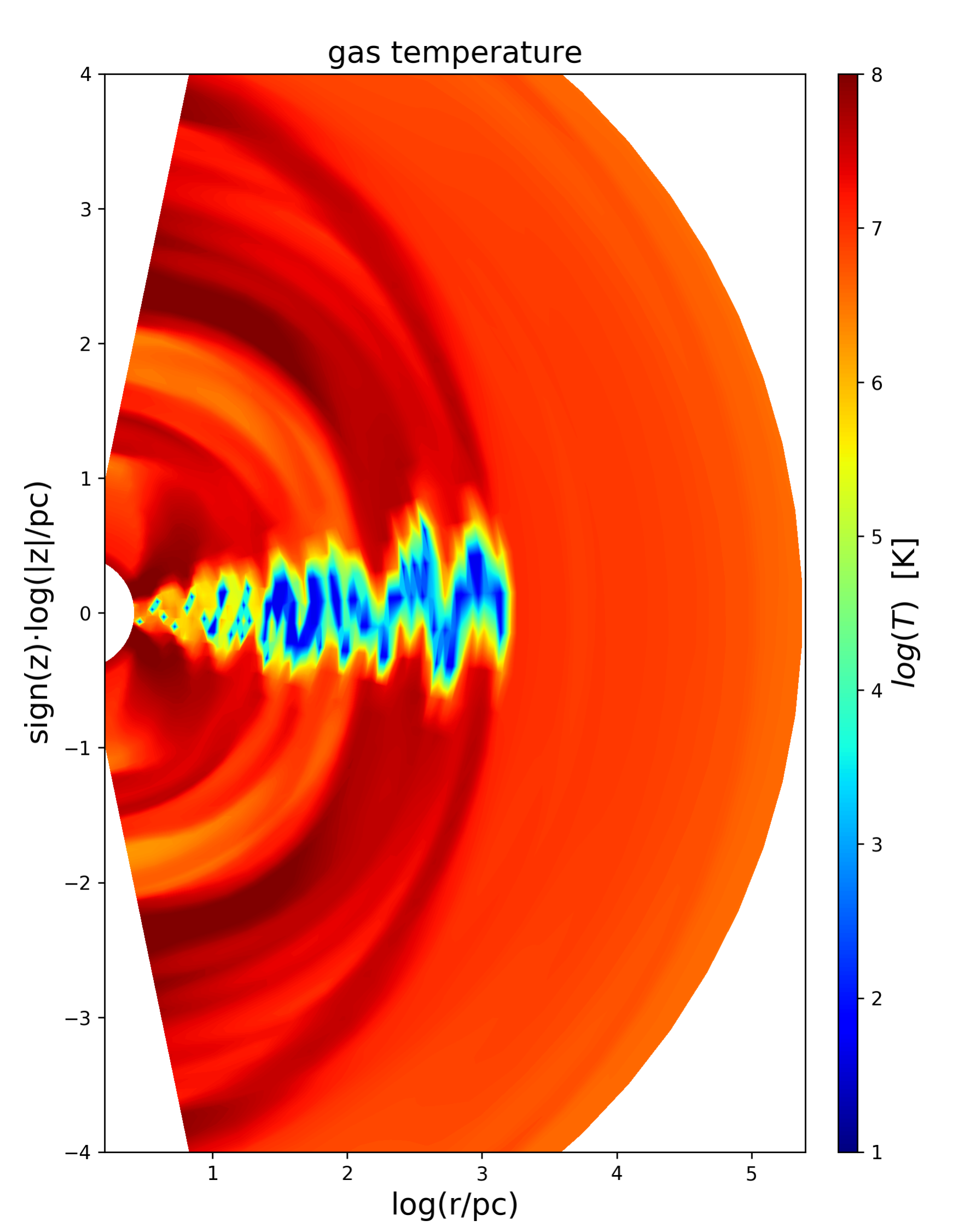}
\includegraphics[width=0.3\textwidth]{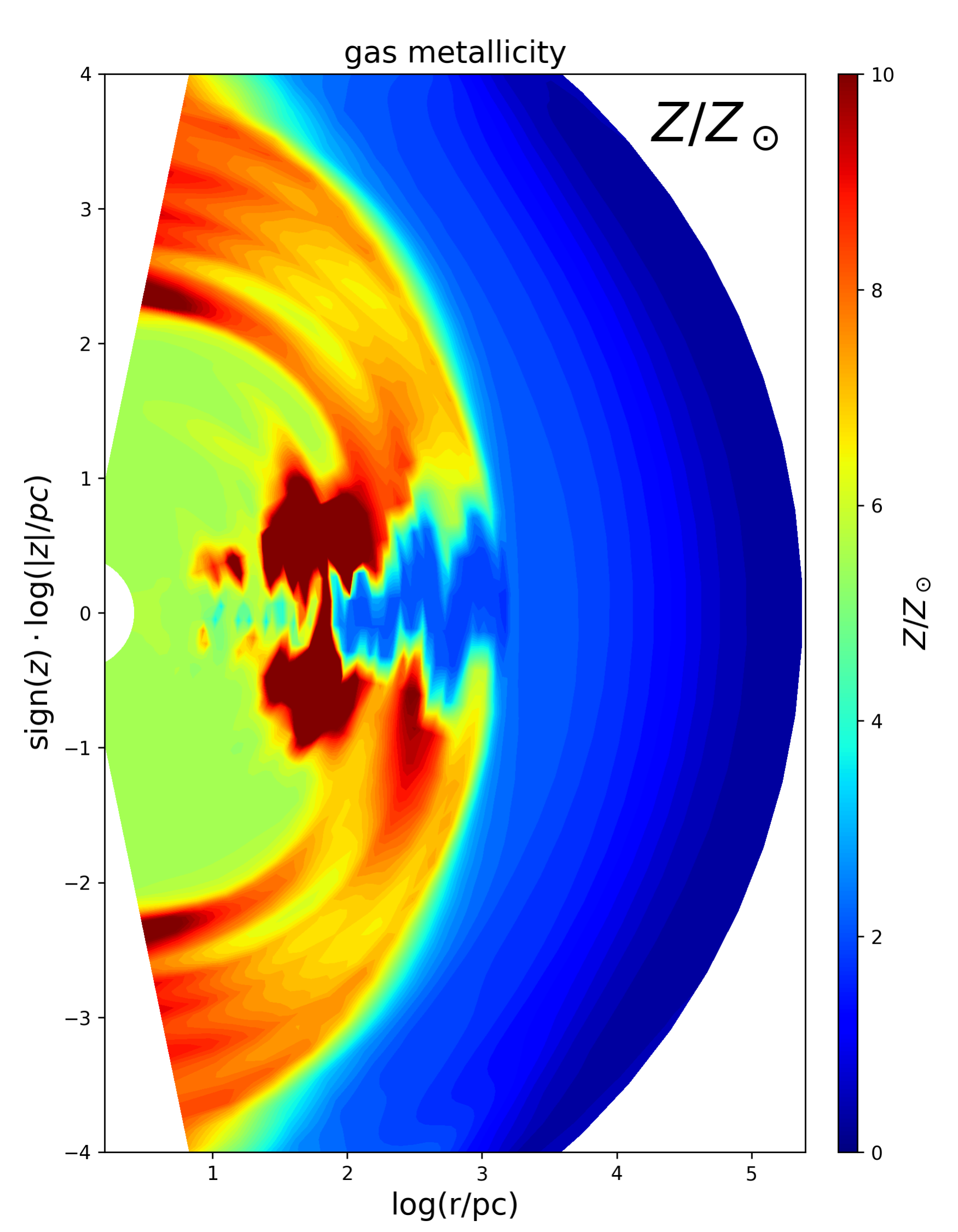} \\
\includegraphics[width=0.3\textwidth]{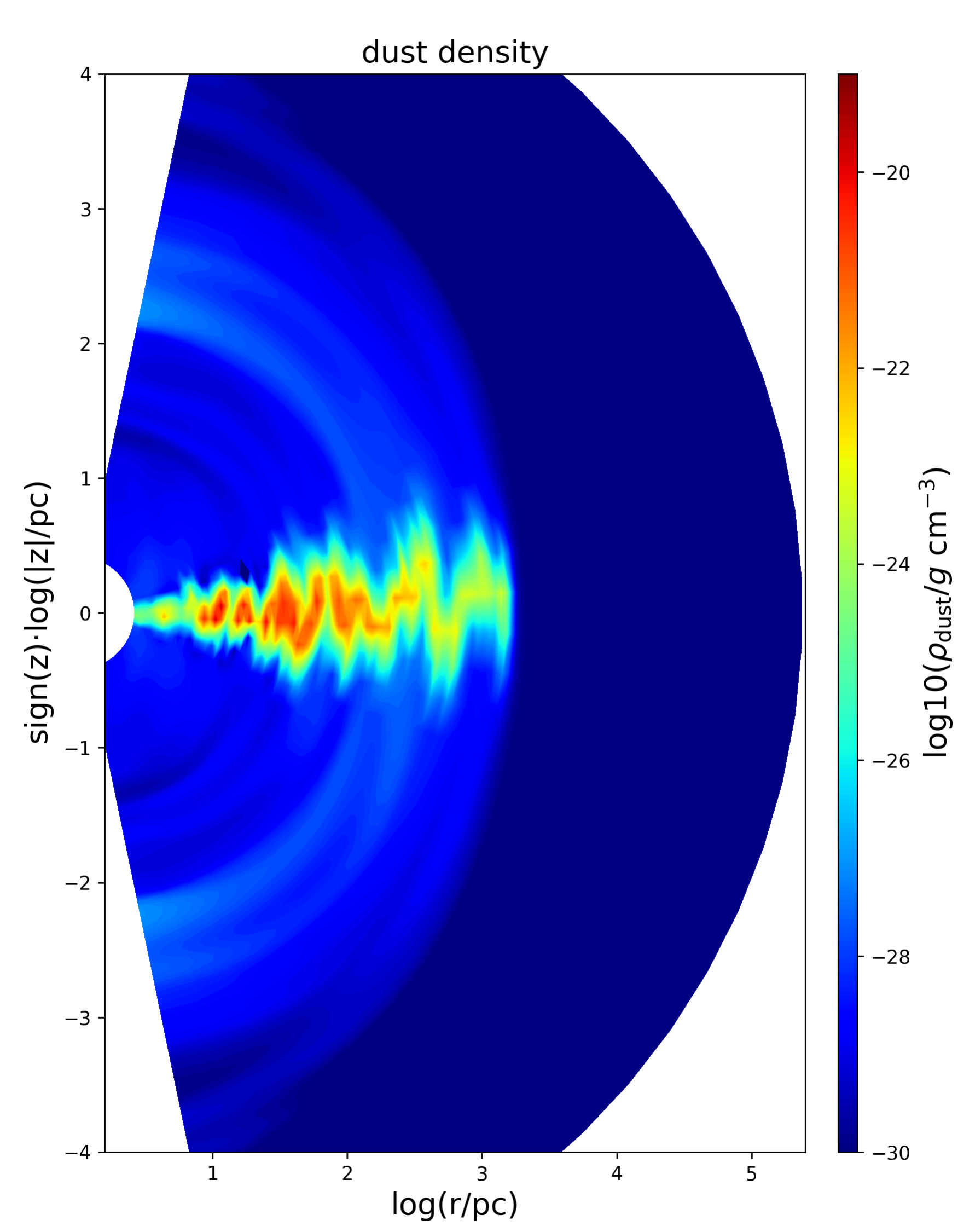}
\includegraphics[width=0.3\textwidth]{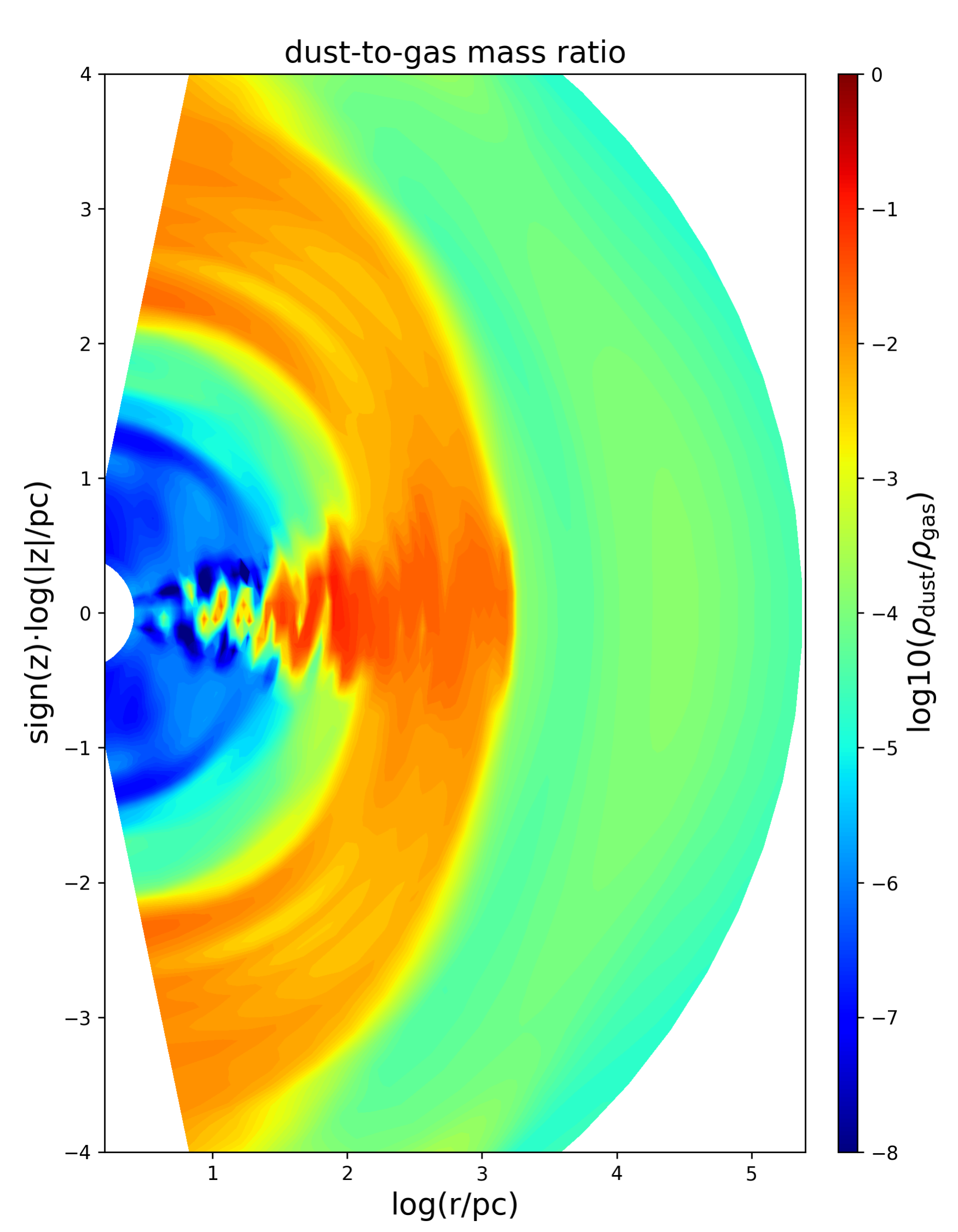}
\includegraphics[width=0.3\textwidth]{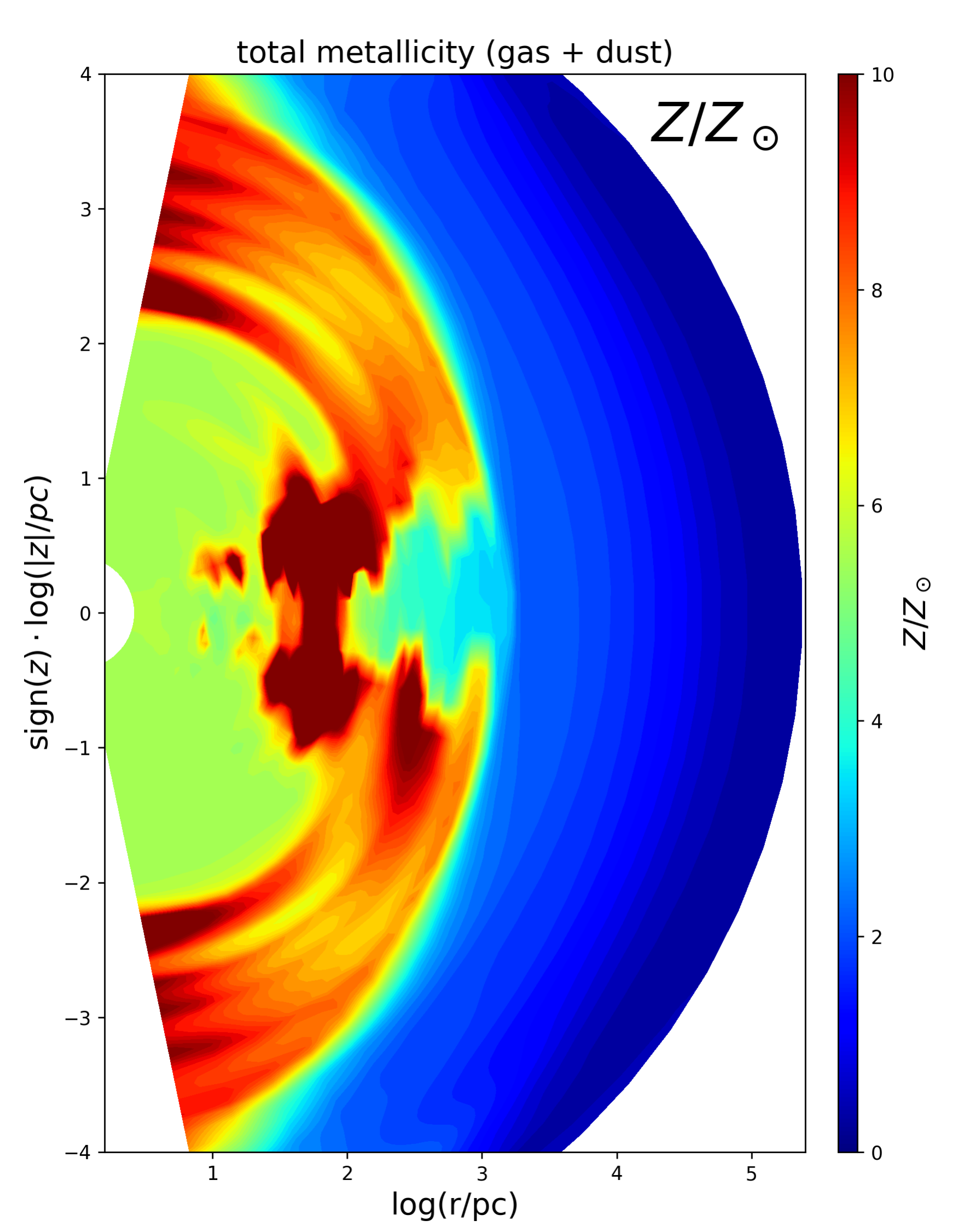}
\caption{Hydrodynamical properties of the cold gas dusty circumnuclear disk at the end of the simulation,
showing the gas density, gas temperature, gas metallicity, dust density, dust-to-gas mass ratio and the total metallicity, respectively.
Note the logarithmic radial scale. }
\label{fig:hydro-properties}
\end{figure*}
\begin{figure*}[htb]
\centering
\includegraphics[width=0.29\textwidth]{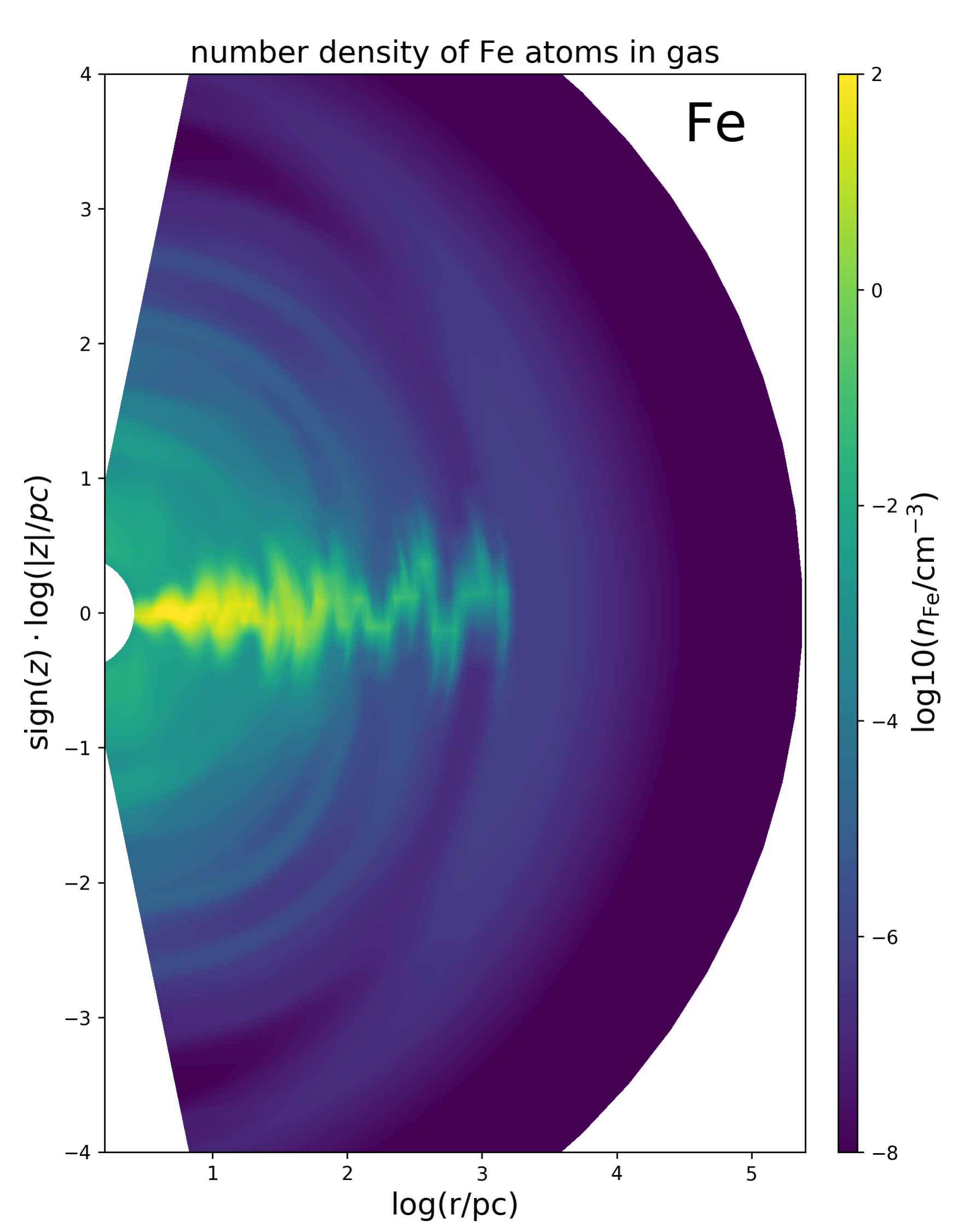}
\includegraphics[width=0.29\textwidth]{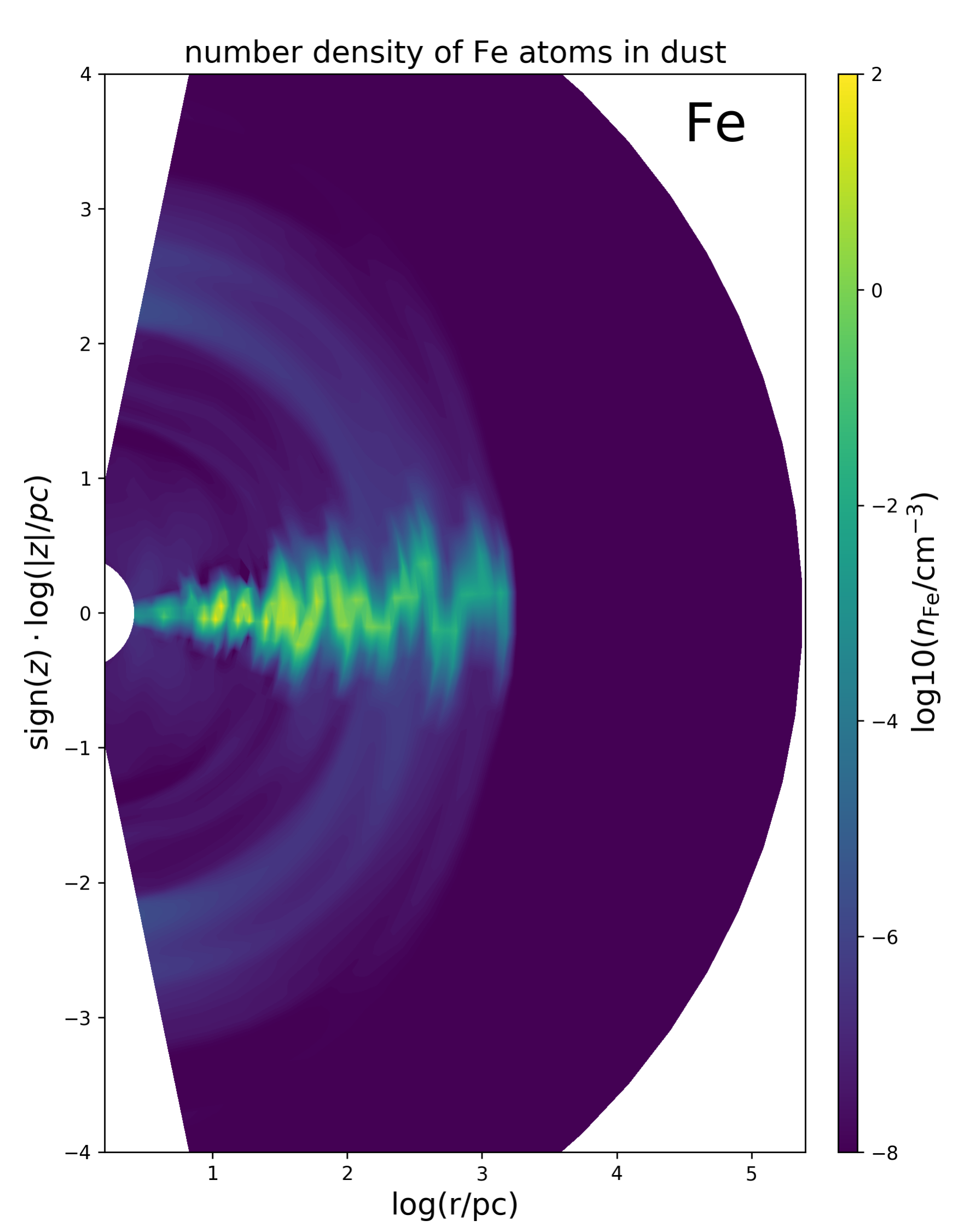}
\includegraphics[width=0.29\textwidth]{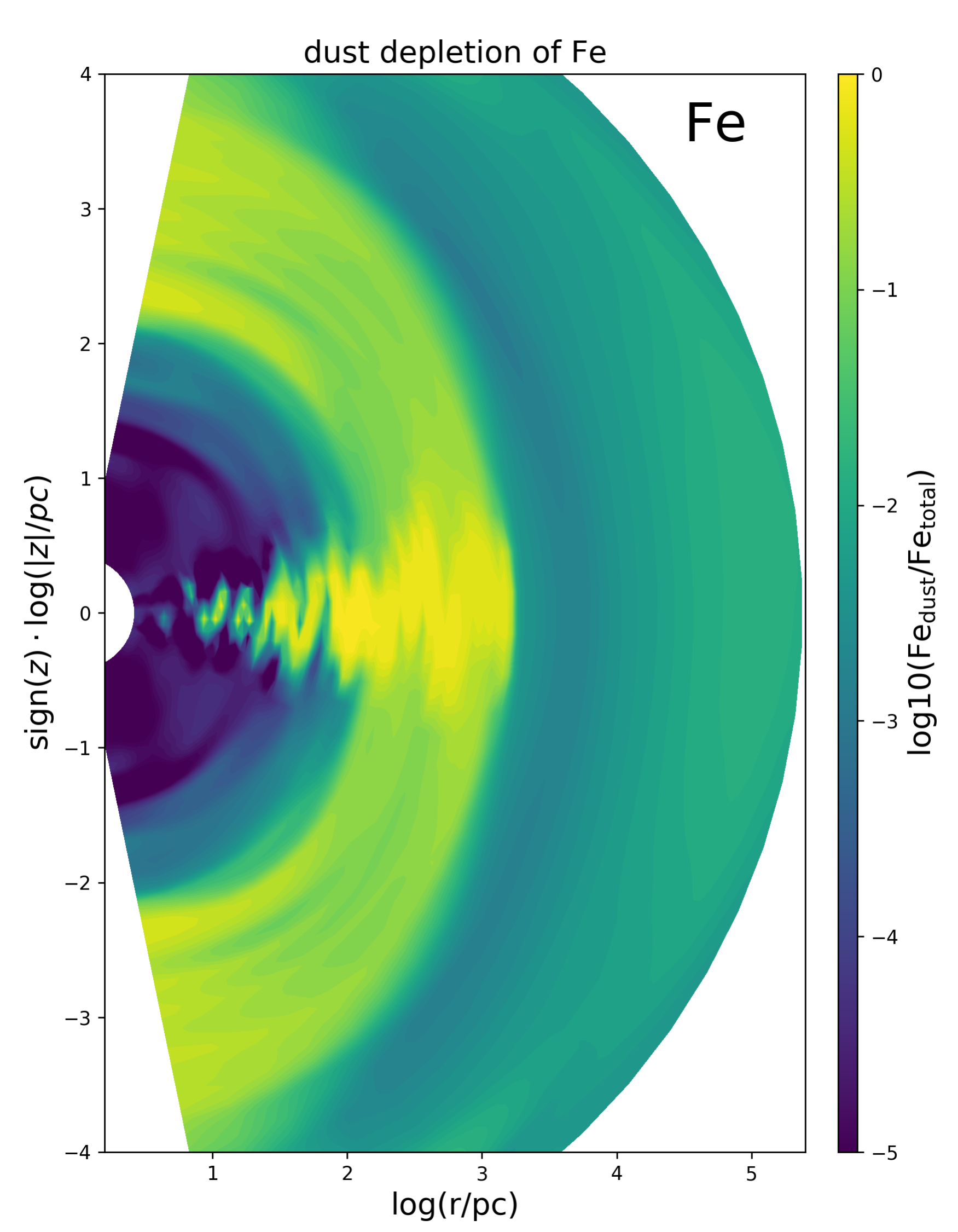}
\caption{Dust distribution and depletion of Fe at the end of the simulation.
Note the logarithmic radial scale.}
\label{fig:dust-depletion-Fe}
\end{figure*}

\section{Results} \label{sec:results}

As in our previous simulations, we found a circumnuclear disk formed in the galaxy center, 
of which the size (half-mass radius) is $\sim 1$ kpc consistent with observations \citep{sarzi_sauron_2006, davis_atlas3d_2011, boizelle_alma_2017}.
\textcolor{black}{The cold gas in the circumnuclear disk is replenished by a normal cooling flow modified by rotation.}
 Most of time, the circumnuclear disk is quiet, 
but once it becomes sufficiently over-dense and self-graviting, 
massive star formation occurs due to the gravitational instability \citep{burkhart_self-gravitating_2019}. 
The latter also permits angular momentum transfer, and thus triggers mass accretion onto the central black hole, 
and therefore induces AGN outbursts after a time lag. Subsequently, the ample energy outputs from the AGN and SNe regulate 
the mass supply onto the circumnuclear disk, stabilize it from the gravitational instability, and force the whole galaxy into a quiescent state 
until another cycle begins (e.g., \citealt{ciotti_cooling_1997, reynolds_inefficient_2015, yang_interplay_2016}).
 
The patterns of the galaxy activities are shown in Figure \ref{fig:sim-overview}, 
in which we plot the history of mass supply to the galaxy center, AGN activities, and star formation. 
Very high time resolution ($\sim10~{\rm yr}$ in our simulation) is needed to compute and to visualize the activity 
since most of the star formation and AGN output occurs in episodes of very short duration.
Note that, while in early starbursts star formation rates can exceed \textcolor{black}{$250M_\odot/{\rm yr}$} 
and in the later ones \textcolor{black}{$\geq20M_\odot/{\rm yr}$}, 
in the quiescent intervals which dominate at late times (the last 1 Gyr in the simulation)
the average star formation rate is less than \textcolor{black}{$1.5 M_\odot/{\rm yr}$} 
\textcolor{black}{(as observed by, e.g., \citealt{kuntschner_sauron_2010, sarzi_atlas3d_2013, davis_atlas3d_2014}; 
see also \citealt{negri_x-ray_2015}).}

In general the galaxy is a prototypical ``quiescent" elliptical galaxy, 
but star formation can exceed that in normal spirals during central outbursts. 
Thus, the top panels of Figure \ref{fig:sim-overview} may present a somewhat misleading visual estimate 
of the fraction of time during which there are starbursts or AGN activity, 
so in the lower panels we show the same quantities averaged over a $10^8$ yr timescale.
In Figure \ref{fig:duty-cycle}, we present the duty cycle (i.e., fraction of time) spent at different levels of AGN activity and star formation. 
It is shown in the left panel of Figure \ref{fig:duty-cycle} that only \textcolor{black}{$\sim$0.5\%} of its lifetime the AGN is in the quasar mode 
(i.e., if the AGN luminosity is larger that 10\% the Eddington limit, $L_{\rm BH}>0.1 L_{\rm Edd}$), 
while during such short duty cycle, the AGN emits \textcolor{black}{$\sim$8\%} of its total energy output in form of radiation.

We can see from the middle panel of Figure \ref{fig:sim-overview} that the instantaneous star formation rate 
is up to $>100M_\odot/{\rm yr}$, while its duty cycle is extremely short (\textcolor{black}{$\lesssim1\times10^{-4}$}; 
see the blue line in the right panel of Figure \ref{fig:duty-cycle}).
The star formation rate is \textcolor{black}{$<2M_\odot/{\rm yr}$} during \textcolor{black}{59\%} of the galaxy lifetime.

In the rest of this section, we focus on the dust, including its hydrodynamical and thermodynamical properties, and its infrared emission. 
In \S\ref{sec:dusty-disk}, we present the spatial distribution of the dust and its effects of metal depletion and obscuration. 
In \S\ref{sec:dust-temperature}, we derive the thermal properties of the dust based on the outputs of our simulation. 
In \S\ref{sec:dust-emission}, we present the infrared emission from dust and compare it with observations.
Finally, we discuss the dependence of our results on the initial conditions and spatial resolution in \S\ref{sec:resolution-study}.

\subsection{Cold Gas Dusty Disk} \label{sec:dusty-disk}

In Figure \ref{fig:hydro-properties}, we show the hydrodynamical properties of the circumnuclear disk, 
including its gas and dust components, at the end of the simulation.
We can see that the gas component of the circumnuclear disk is dense and cold, 
which is ideal for star formation and dust grain growth. 
In the upper-mid panel, we see that in most of the volume the gas is virialized, with a typical temperature around 1 keV. 
In the circumnuclear disk, the gas can be cooled far below $10^4$ K until it hits the temperature floor of $50$ K, 
as we now allow for dust cooling of the hot gas. In the lower-left panel, we plot the dust density. 
In the lower-mid panel, we present the dust-to-gas mass ratio, 
which is up to \textcolor{black}{$\sim 0.1$}, and it is much higher than the typical value in the solar neighborhood ---
it indicates that the dust depletion of metals can be very significant,
though in the innermost part of the disk, AGN outbursts have efficiently destroyed all dust \citep{kawakatu_obscuring_2019}.

To demonstrate the effects of dust depletion of metals, in Figure \ref{fig:hydro-properties} we plot the gas metallicity 
and total (gas and dust) metallicity in the upper-right and lower-right panels, respectively. 
We can see that the metallicities are  \textcolor{black}{significantly enhanced in the inner region} 
but there is a void in the spatial distribution of the gas metallicity in the outer disk, 
which is coincident with the peak of the dust-to-gas mass ratio (see the lower-mid panel in Figure \ref{fig:hydro-properties}).
\textcolor{black}{In the bulk of the hot x-ray emitting gas, the metallicity is below the solar value in average in the late times, 
but the BAL winds can have a highly super-solar metal abundance. 
The innermost and outermost regions are nearly dust free,
the former because of AGN activity, the latter due to the low metallicity of cosmological inflow gas.}
We note that the circumnuclear disk is dusty, especially in its outer region where most of the metals are in dust grains.
In Figure \ref{fig:dust-depletion-Fe}, we present the mass distribution of Fe in forms of gas and dust grains. 
We can see that the mass density of the gas-phase Fe peaks in the inner region of the circumnuclear disk, 
and it is spread in the polar regions due to the AGN winds. 
From the right panel of Figure \ref{fig:dust-depletion-Fe}, we can see the significant dust depletion of Fe; 
almost all the Fe is in dust grains in the outer disk.

\begin{figure}[htb]
\centering
\includegraphics[width=0.475\textwidth]{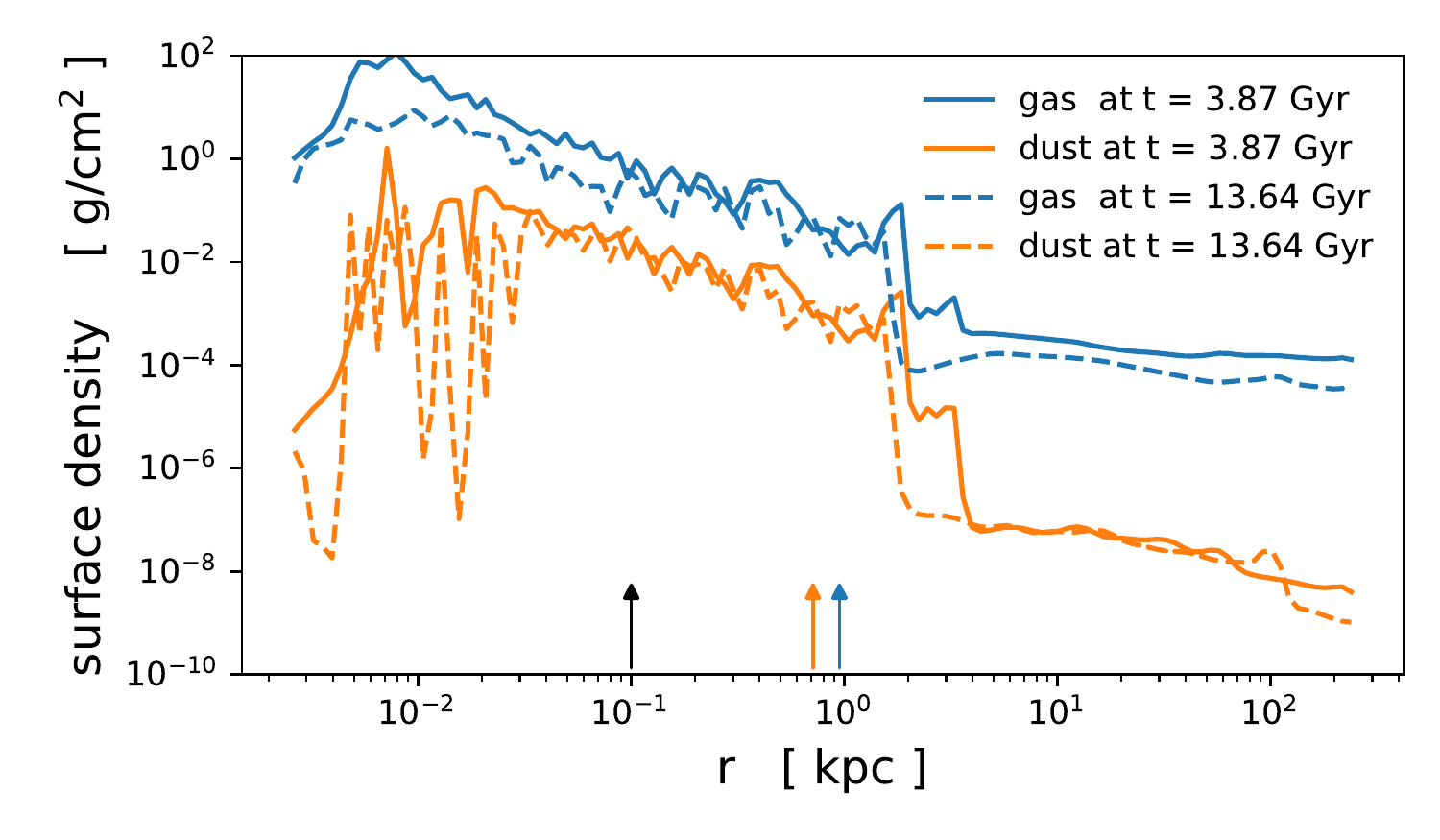}
\caption{Surface density of the gas and dust component in the circumnuclear disk at the end of the simulation.
The vertical black arrow indicates the radius of influence of the central black hole within which it dominates the gravitational potential;
The vertical orange and blue arrows indicate the half-mass radius of the dust disk and that of the cold/warm gas ($T_{e}<10^5$ K) disk, respectively.}
\label{fig:surface-density}
\end{figure}

In Figure \ref{fig:surface-density} we plot the surface density of the gas- and dust-components with blue and orange lines, respectively. 
The solid lines represent the results during an outburst \textcolor{black}{at $t=3.87$ Gyr}, 
while the dashed lines are the results (\textcolor{black}{at $t=13.64$ Gyr}) when the system is relatively quiescent. 
We can see from the figures above that the spatial distribution of dust overlaps with that of the gas component, but unlike the gas component, 
the dust density peaks in the outer disk rather than in the inner disk, 
as the gas temperature is relatively high in the inner disk due to the AGN feedback and supernova heating, 
which makes it difficult for the dust grains to survive.

\begin{figure}[htb]
\centering
\includegraphics[width=0.475\textwidth]{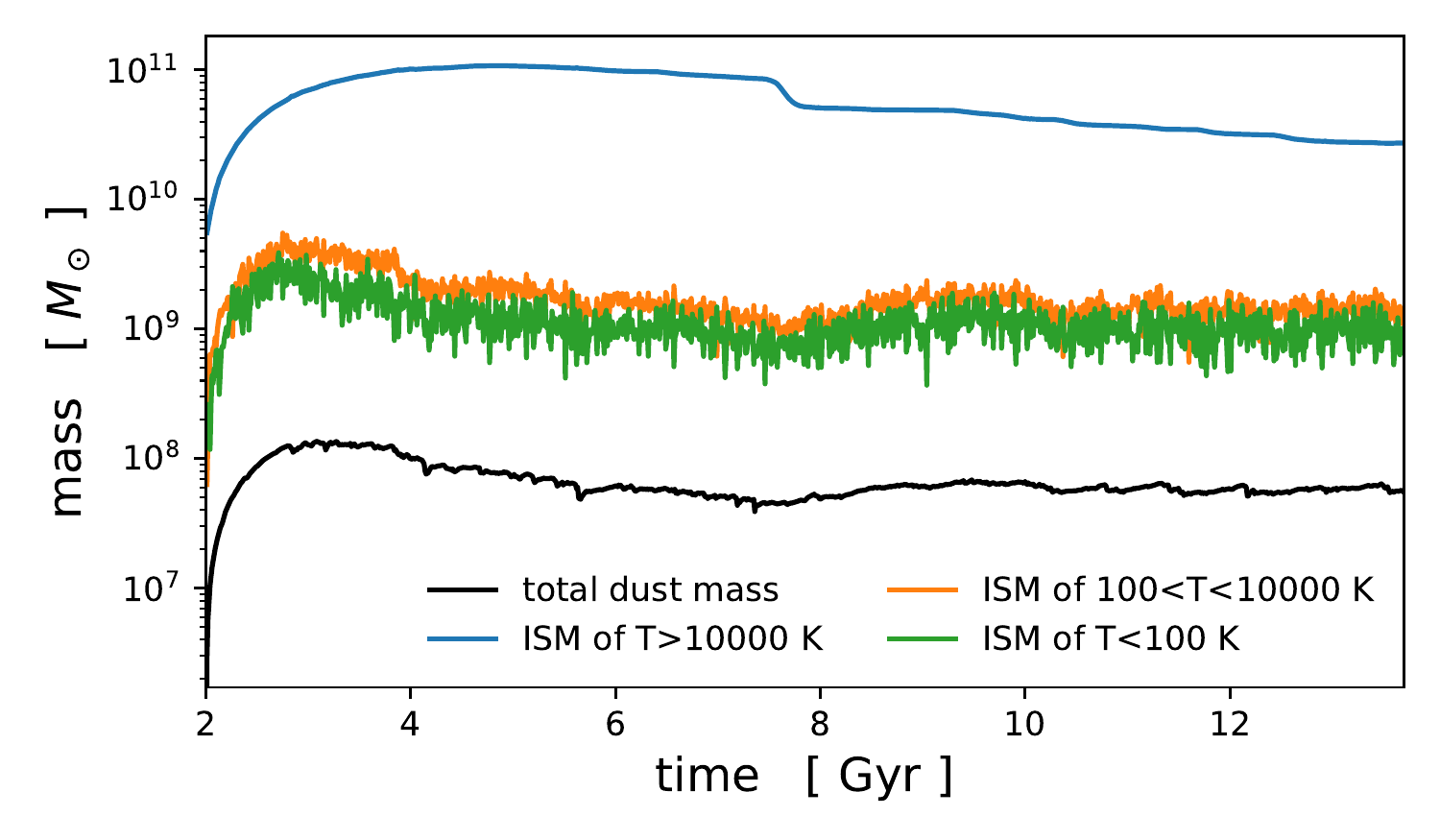}
\caption{\textcolor{black}{Total mass of dust grains and gaseous ISM in different range of temperature. 
As expected, hot X-ray emitting gas dominates.}}
\label{fig:ism-mass-hst}
\end{figure}

In Figure \ref{fig:ism-mass-hst} we plot the total mass of dust grains and gaseous ISM in different range of temperature. 
It is shown that the total dust mass is typically $5\times10^7M_\odot$, the typical masses of the cold circumnuclear
disk and the X-ray emitting hot ISM are $\sim10^9$ and \textcolor{black}{$10^{11}M_\odot$}, respectively.

Dust obscuration of photons is known to be very important in many astrophysical systems,
especially at UV and optical wavelengths, the dust opacity in AGN environments can dominate over electron scattering.
In next subsection, we calculate the infrared emission from the dust, where the irradiation by the AGN and starlight is
an important energy source for the dust emission. Therefore, it is useful to discuss below the dust obscuration of 
both the AGN radiation and the starlight in/through the circumnuclear disk.

\begin{figure}[htb]
\centering
\includegraphics[width=0.46\textwidth]{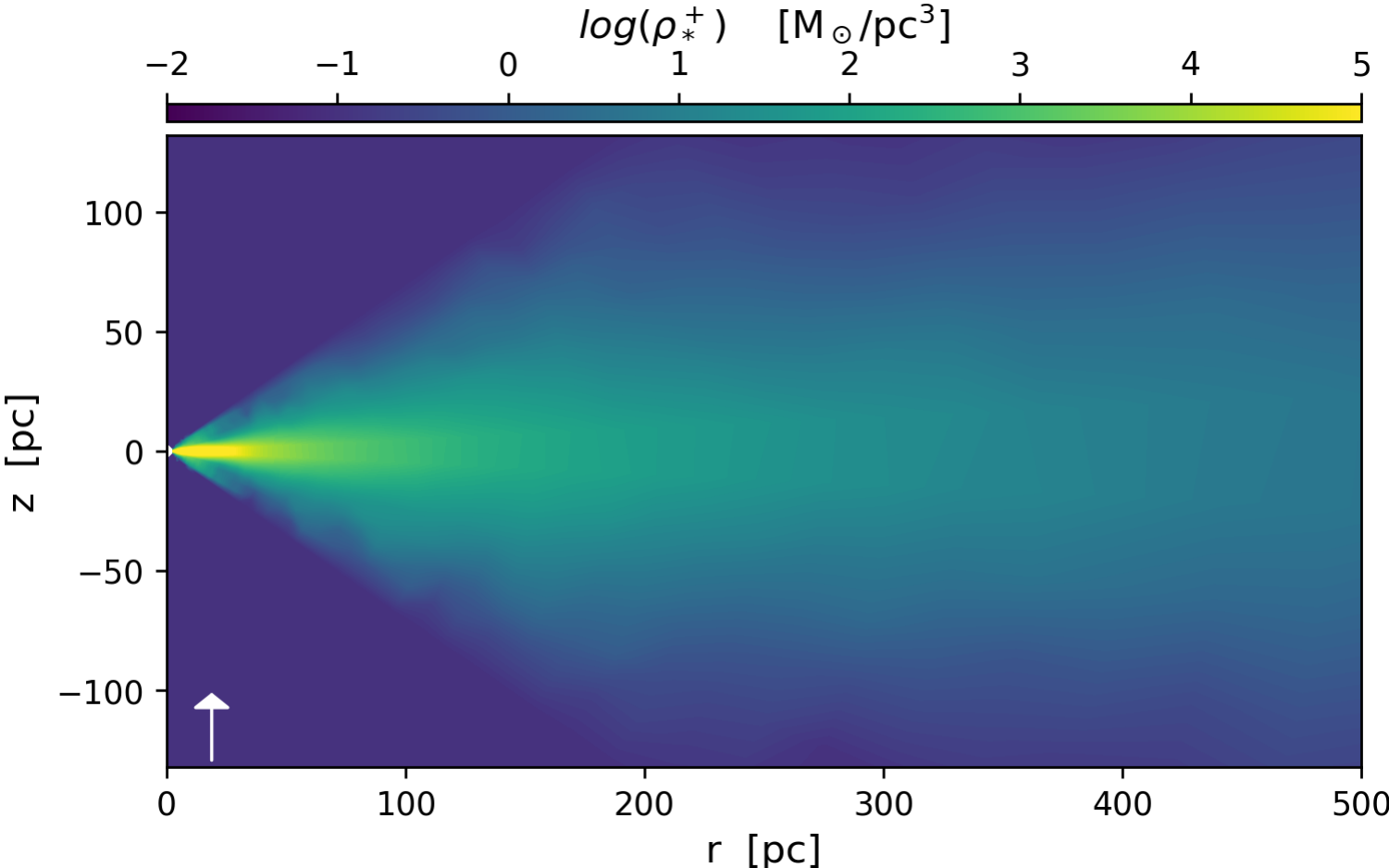}
\caption{The cumulative star formation \textcolor{black}{(stellar mass density)} in the circumnuclear disk at the end of the simulation. 
The half-mass radius is $\sim 20$ parsec as indicated by the vertical arrow.}
\label{fig:cumulative-star-formation}
\end{figure}

Before we evaluate the dust absorption of starlight,  
we present in Figure \ref{fig:cumulative-star-formation} the cumulative star formation during the simulation.
We can see that almost all of the star formation ( \textcolor{black}{$\sim3.3\times10^{10} M_\odot$}) is 
embedded in in the dusty disk, where the gas is over-dense. 
The characteristic radius of the star forming region is only $\sim20$ parsec 
and would be difficult to detect except in the nearest elliptical systems.
\textcolor{black}{Noted that: i)} in our star formation criterion (\S\ref{sec:jeans-criterion}), we have not yet included the shear
and turbulent velocity dispersion {\it within} cells. Allowing for this would reduce star formation in the innermost regions;
\textcolor{black}{ii) we do not include the dynamical and structural effects of the new stars, which may dominate the stellar mass and 
shift the dynamics significantly in the innermost region.}
One can expect that the new stars are an important source of UV photons which can be easily absorbed by dust grains,
and thus can be effective in heating up the dust, and enabling the latter to emit in infrared.

\begin{figure}[htb]
\centering
\includegraphics[width=0.45\textwidth]{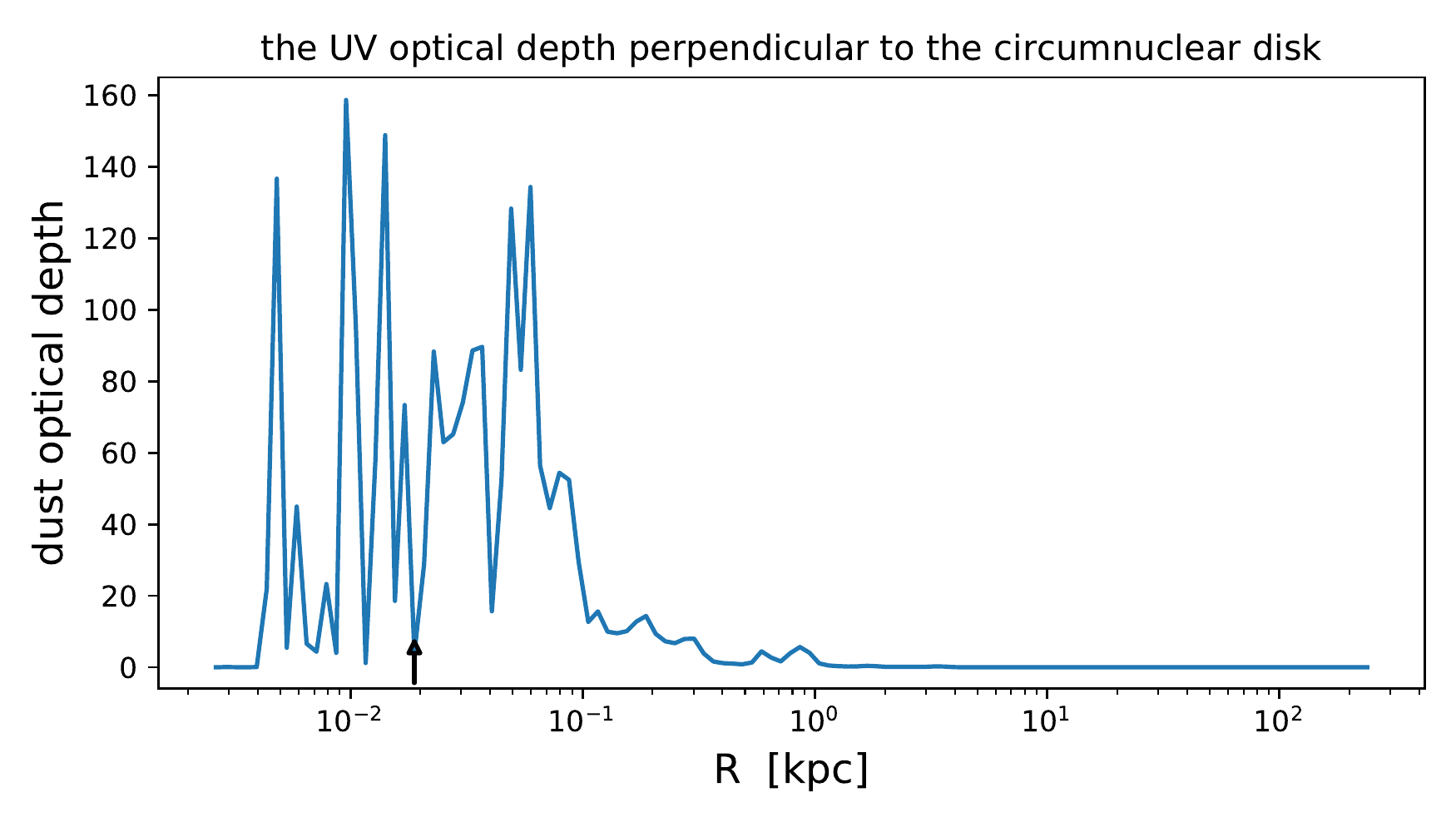}
\caption{Dust optical depth in the UV band perpendicular to the circumnuclear disk  
during the starburst \textcolor{black}{at $t_{\rm age}=4.13$ Gyr}. 
The vertical arrow is at the half-mass radius of the new stars.}
\label{fig:dust-OpticalDepth_UV_newStar}
\end{figure}

\begin{figure}[htb]
\centering
\includegraphics[width=0.4\textwidth]{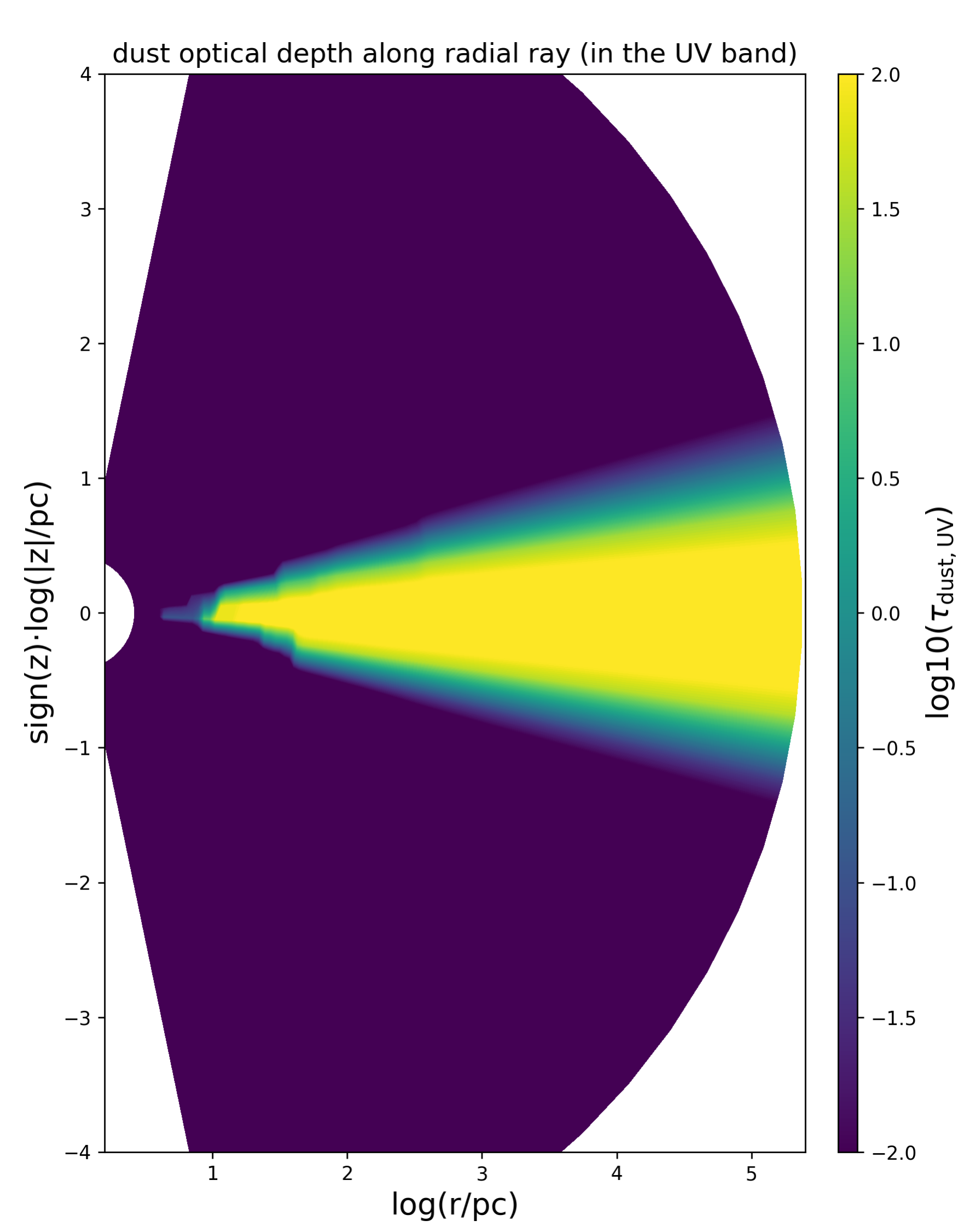}
\caption{Dust optical depth at the end of the simulation in the UV band along radial directions from the center.  
Note the logarithmic radial scale. }
\label{fig:dust-OpticalDepth_UV_AGN}
\end{figure}

In Figure \ref{fig:dust-OpticalDepth_UV_newStar}, we plot the dust optical depth 
for the UV photons emitted in the circumnuclear disk during the starburst \textcolor{black}{at $t_{\rm age}=4.13$ Gyr}.
For simplicity, the optical depth is integrated along the $\theta-$ direction.  
Again, we can see that the dusty disk is optically thick to the star light, i.e., most of the stellar radiation is expected to be 
absorbed within the disk, and only the ``runaway stars'' would be visible to distant observers.
Of course, after the outbursts when the dust clears, 
an A type spectrum would be seen from stars distributed as shown in Figure  \ref{fig:cumulative-star-formation}.

In Figure \ref{fig:dust-OpticalDepth_UV_AGN}, we plot the dust optical depth for the UV photons from the AGN 
(see Equation \ref{eq:dust-opt-depth-uv}) at redshift zero, where the UV optical depth is integrated along radial directions with fixed $\theta$ since the AGN can be 
treated as a point source. 
Comparing to the lower-left panel of Figure \ref{fig:hydro-properties}, we find that most of the dusty disk 
is optically thick to the UV photons from the central AGN, i.e., all the AGN emission is obscured behind the circumnuclear disk. 
The solid angle of the dusty disk is \textcolor{black}{$\sim2.36$ steradians}, i.e., the covering factor is \textcolor{black}{$\sim19\%$} of the $4\pi$ solid angle.

\subsection{Dust Thermodynamics} \label{sec:dust-temperature}
In this section, we describe how we derive the infrared emission of dust from the simulation outputs. 
Following \citet{hensley_grain_2014}, we assume the dust emission in infrared 
is in equilibrium with its heating processes, i.e., 
\begin{equation}  \label{eq:dust-heating}
 \begin{aligned}
   P^{\rm lR}_{\rm dust}  &\equiv \sigma T_{\rm dust}^4 
                                         \int_{a_{\mathrm{min}}}^{a_{\mathrm{max}}} \! {d}a \frac{{d}n_d}{{d}a}  
                                         \cdot 4 \pi a^2 \bar{Q}(a,T_{\rm dust}) \\
                   &= H^{\rm rad}_{\rm dust,abs} +H^{\rm gas}_{\rm dust,collision}             
 \end{aligned}
\end{equation}
where $P^{\rm lR}_{\rm dust}$ is the dust emission power per unit volume. 
$T_{\rm dust}$ is the dust temperature, $\bar{Q}(a,T_{\rm dust})$ is 
the Planck-averaged emission efficiency (cf. \citealt{hensley_grain_2014}, their Equation 33,
\textcolor{black}{where the Planck averaged absorption cross sections for silicate grains are used; 
see also \citealt{schurer_modelling_2009}; \citealt{draine_physics_2011}}).
Therefore, provided the heating terms, the dust temperature can be calculated according to the equation above.


The heating sources are mainly contributed by collisional heating $H^{\rm gas}_{\rm dust,collision}$ 
and radiative heating $H^{\rm rad}_{\rm dust,abs}$. 
The collisional heating is calculated according to Equation \ref{eq:dust-collisional-heating}, 
and assumes equilibrium when the cooling timescale is short as in the hydrodynamical part of the simulation. 
Unfortunately, the assessment of the radiative heating of dust is much more complicated --- 
there are two radiative heating sources for the dust grains, i.e., the AGN radiation and the stellar radiation, 
\begin{equation} 
    H^{\rm rad}_{\rm dust,abs} = H^{\rm AGN}_{\rm dust,abs}  + H^{\rm stars}_{\rm dust,abs}.
\end{equation}

\begin{figure*}[htb]
\centering
\includegraphics[width=0.75\textwidth]{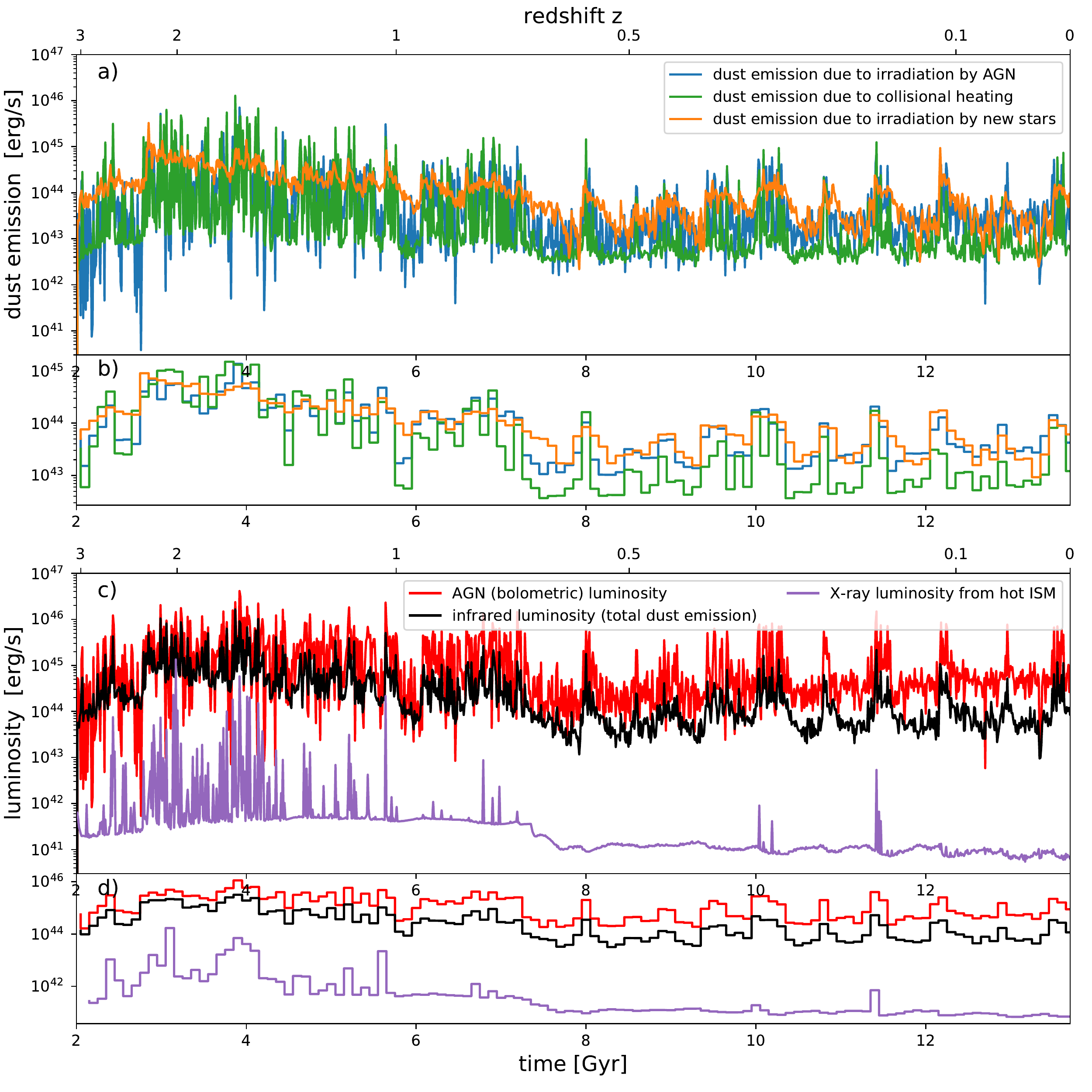}
\caption{Infrared emission from dust. 
\textcolor{black}{A rough estimate of the wavelength of the infrared emission can be found 
by considering Figure \ref{fig:IR-half-radius} (top panel), which shows the strong correlation between dust temperature and luminosity.}
\textcolor{black}{The time resolution in Panels (a, c) is $\sim10^7~{\rm yr}$, while Panels (b, d) average over a $10^8~{\rm yr}$ timescale.}
\\ 
}
\label{fig:dust-IR-emission}
\centering
\includegraphics[width=0.3\textwidth]{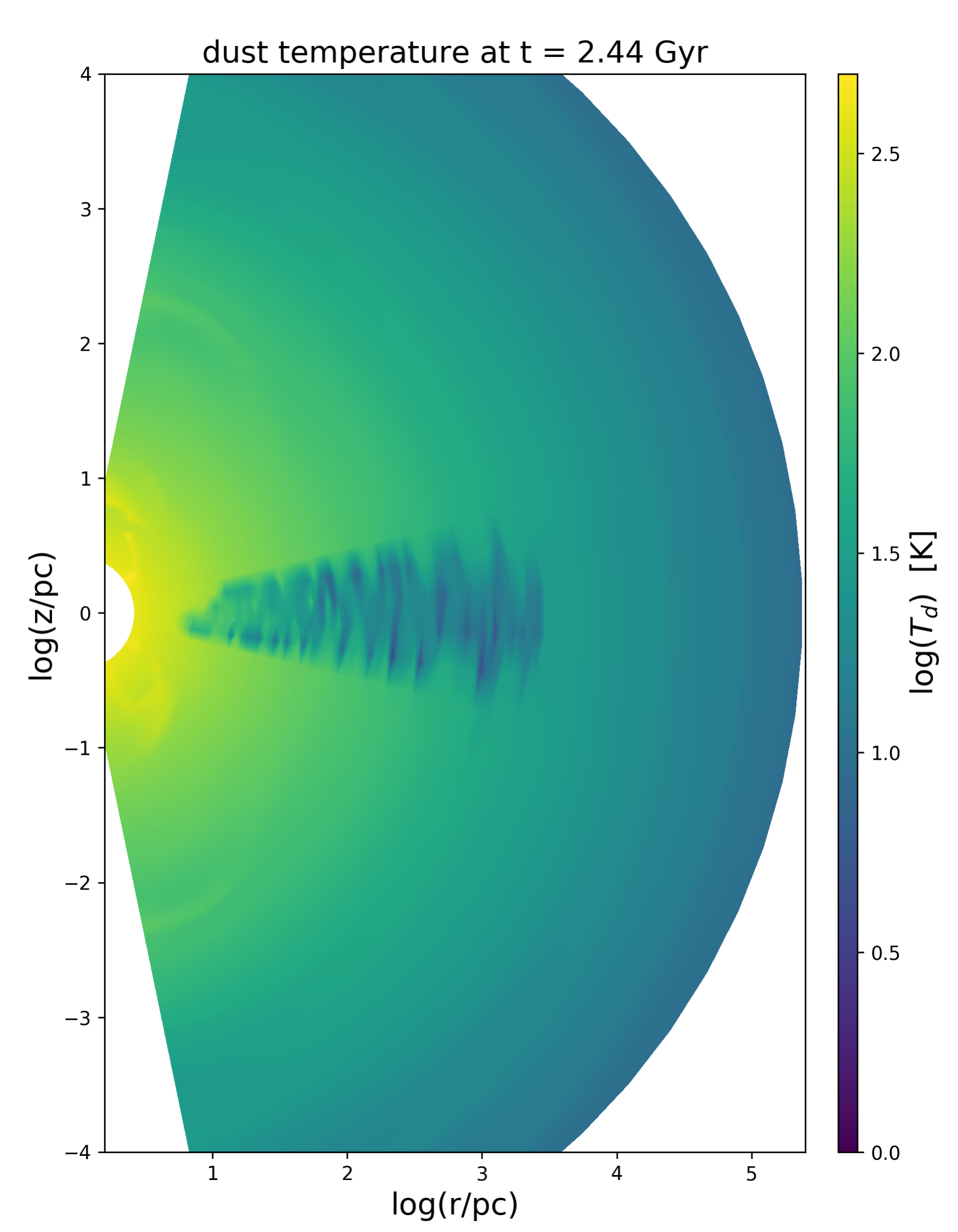}
\includegraphics[width=0.3\textwidth]{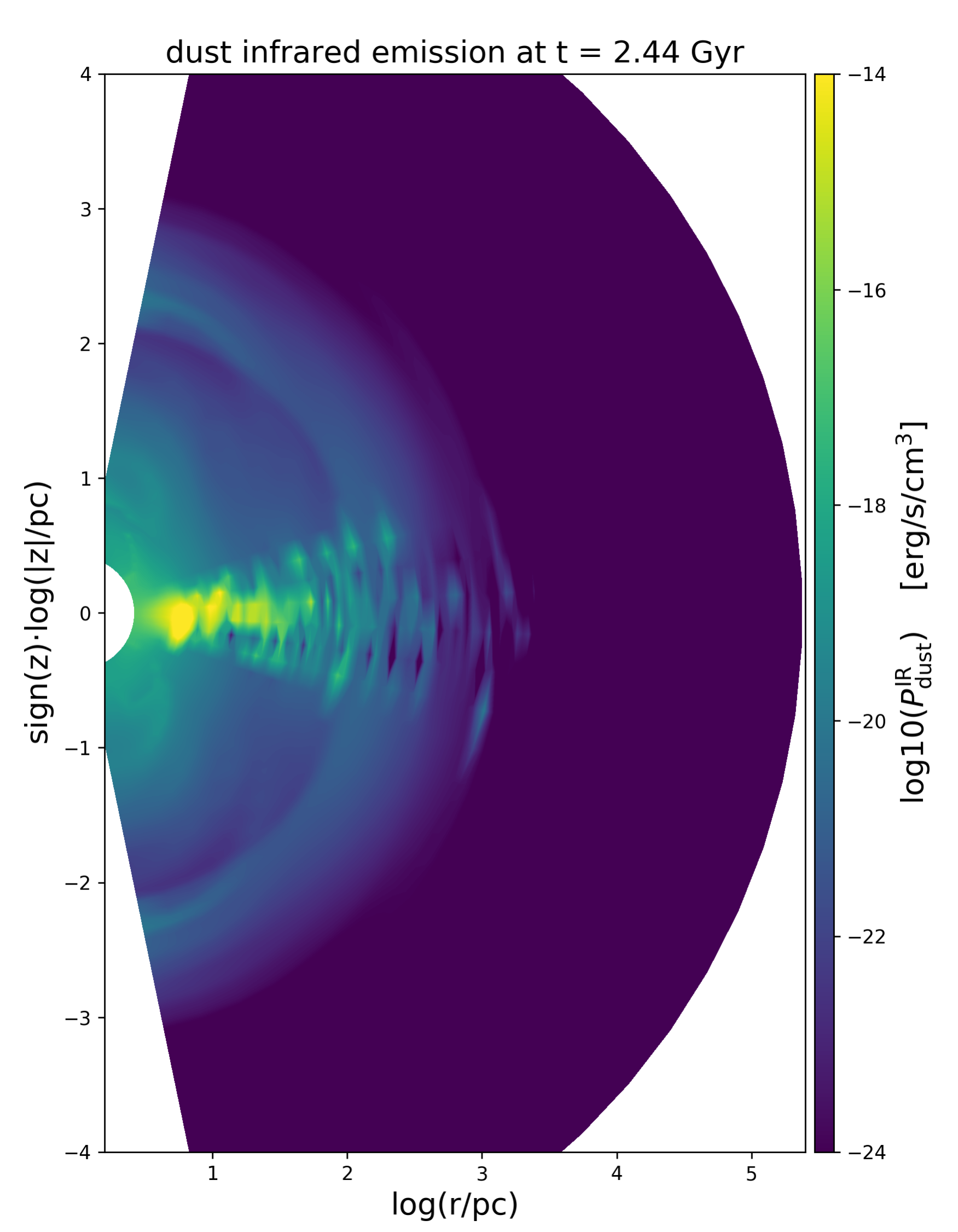}
\includegraphics[width=0.3\textwidth]{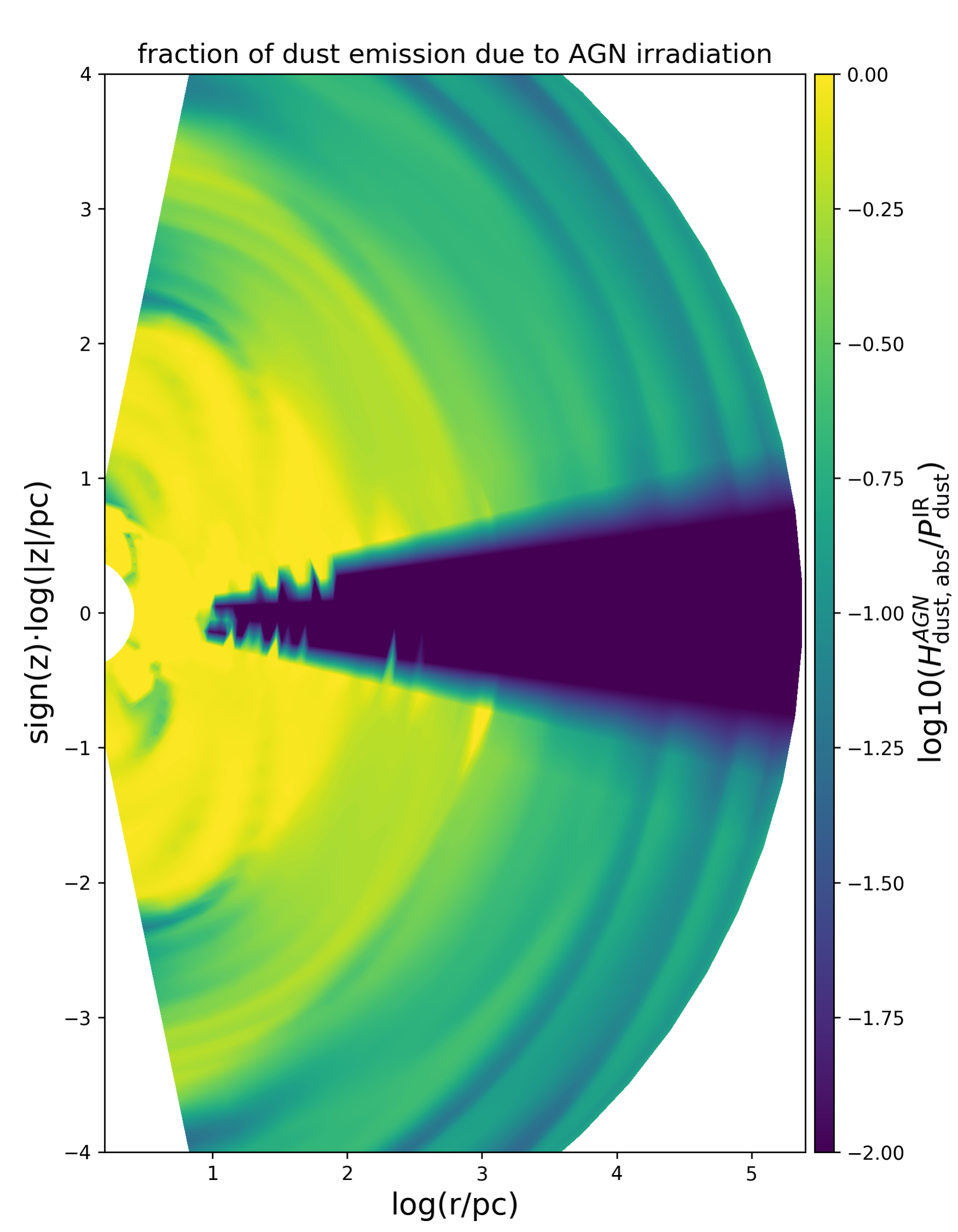}
\caption{Dust temperature (left panel), its infrared emission power per unit volume  (middle panel) 
and the fraction of the dust emission due to AGN irradiation (right panel) 
during the outburst \textcolor{black}{at $t=2.44$} Gyr, when the AGN output is \textcolor{black}{$0.08 L_{\rm edd}$}, 
the star formation rate is \textcolor{black}{$1.7M_\odot/{\rm yr}$}, 
and the total infrared luminosity \textcolor{black}{$L_{\rm TIR}=7.4\times10^{44}$ erg/s}.
Note the logarithmic radial scale.}
\label{fig:dust-IR-emissivity}
\end{figure*}

\begin{figure}[htb]
\centering
\includegraphics[width=0.475\textwidth]{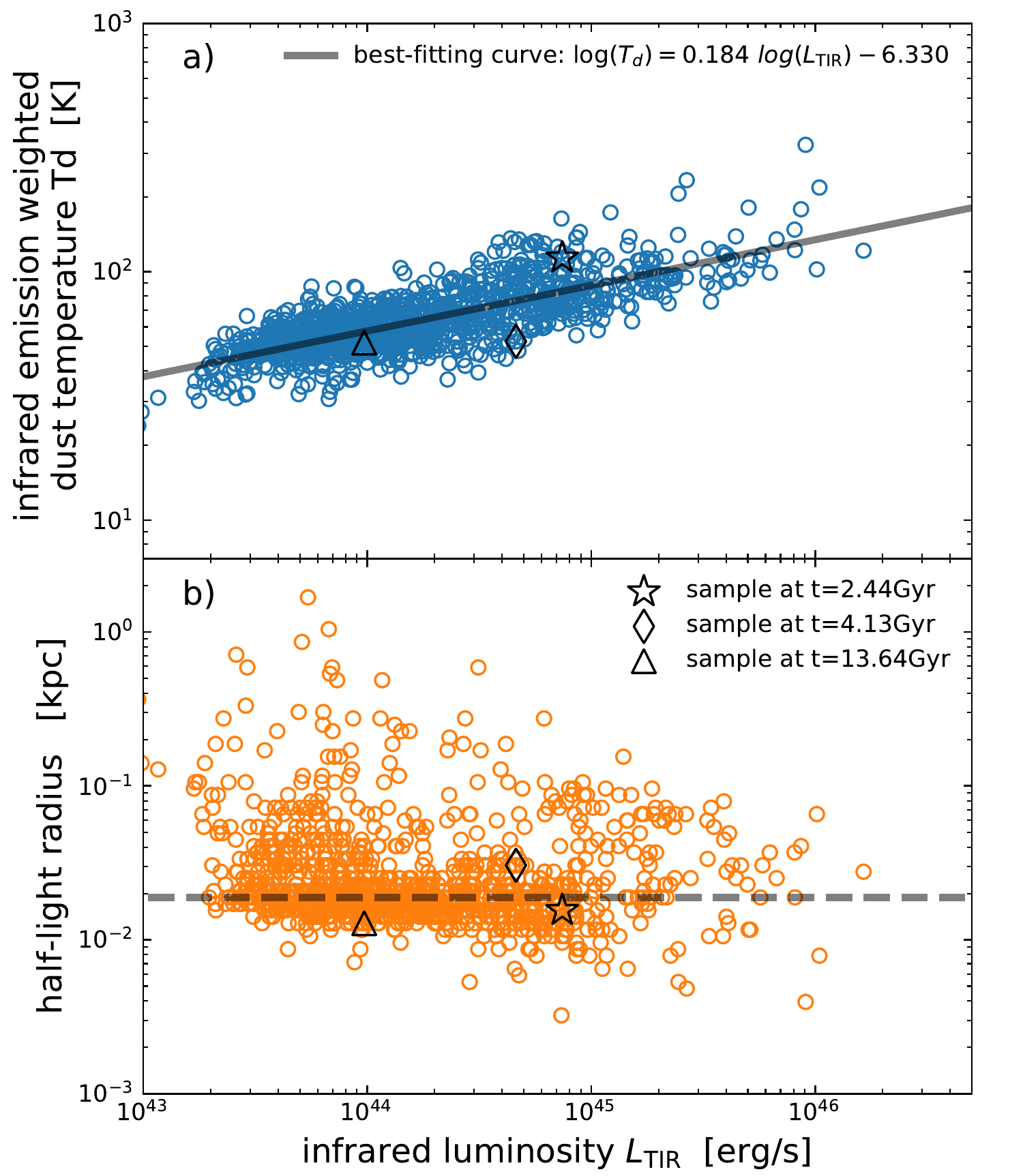}
\caption{Top panel -- infrared emission weighted dust temperature. 
\textcolor{black}{We can see a strong correlation between dust temperature and luminosity shown in the figure, 
which also makes it possible to have a rough estimate of the wavelength of the infrared emission}
\textcolor{black}{The empty circles are the data sampled for every 10 Myr throughout the whole simulation.}
The black solid line is our best fitting curve;
Bottom panel -- half-light radius of the dust infrared emission. 
The dashed line shows the median size of 20 parsec. 
}
\label{fig:IR-half-radius}
\end{figure}

\noindent
The dust heating by AGN irradiation $H^{\rm AGN}_{\rm dust,abs}$ is calculated 
similarly to Equation \ref{eq:dust-radiative-heating} \& \ref{eq:dust-opt-depth-uv}, 
but with both the UV and optical photons. 
However, one can not assess $H^{\rm stars}_{\rm dust,abs}$ self-consistently 
without including sophisticated radiative transfer, and the latter is extremely computationally expensive. 
Fortunately, we find the dust is usually optically thick to the local stellar radiation of the new stars, 
which makes it possible to allow some simplification, 
i.e., we estimate  $H^{\rm stars}_{\rm dust,abs}$ directly with the local stellar radiation 
from the new stars $\dot{E}^{+}_{\rm \star,rad}$, 
\textcolor{black}{assuming that all the local stellar radiation is absorbed by dust.}
\begin{equation} \label{eq:ism-radiative-heating}
     H^{\rm stars}_{\rm dust,abs} \sim \dot{E}^{+}_{\rm \star,rad} \sim \dot{E}^{+}_{\rm \star,UV} + \dot{E}^{+}_{\rm \star,opt},
\end{equation}
where the stellar radiation of the new stars is mainly in the UV ($\dot{E}^{+}_{\rm \star,UV}$) 
and optical ($\dot{E}^{+}_{\rm \star,opt}$) bands.

Following \citet{ciotti_agn_2012}, we compute the UV and optical luminosity per unit volume of the new stars 
according to the star formation history, 
\begin{equation} 
	\dot{E}^{+}_{\rm \star,UV} =  \frac{\epsilon_{\rm UV} c^2}{t_{\rm UV}}
                  \int^t_0 \dot{\rho}^+_\star (t^\prime) 
                  \cdot e^{-\frac{t-t^\prime}{t_{\rm UV}}} dt^\prime,
\end{equation}
\begin{equation} 
	\dot{E}^{+}_{\rm \star,opt} =  \frac{\epsilon_{\rm opt} c^2}{t_{\rm opt}}
                  \int^t_0 \dot{\rho}^+_\star (t^\prime) 
                  \cdot e^{-\frac{t-t^\prime}{t_{\rm opt}}} dt^\prime,
\end{equation}
where $\dot{\rho}^+_\star$ is the star formation rate density (which is given by the \texttt{MACER} simulations). 
In the equations above, $\epsilon_{\rm opt}=1.24\times10^{-3}$, $\epsilon_{\rm UV}=8.65\times10^{-5}$, 
$t_{\rm opt}=1.54\times10^8$ year, $t_{\rm UV}=2.56\times10^6$ year 
are the efficiency and characteristic time of optical and UV emission, respectively.

\textcolor{black}{Note that we ignore the dust opacity of the infrared emission itself in the calculations above, 
which could be important in near or mid-near infrared band \citep{honig_redefining_2019}, and we leave it for future work.}

\subsection{Infrared Emission} \label{sec:dust-emission}

In Figure \ref{fig:dust-IR-emission}, we present the results from the data post-processing of the dust infrared emission. 
In the calculation, we assume the gas and dust are transparent to the infrared photons, 
therefore, the total infrared luminosity \textcolor{black}{$L_{\rm TIR}$} of the galaxy is simply the sum of the dust infrared emission in each cell,
and it makes it possible to assess the contributions of dust emission by the AGN irradiation, star light, and dust-gas
collisional heating, separately (in the upper panels (a, b) in Figure \ref{fig:dust-IR-emission}).  
In the lower panels (c, d), we present the total infrared luminosity (black line).


\begin{figure}[htb]
\centering
\includegraphics[width=0.475\textwidth]{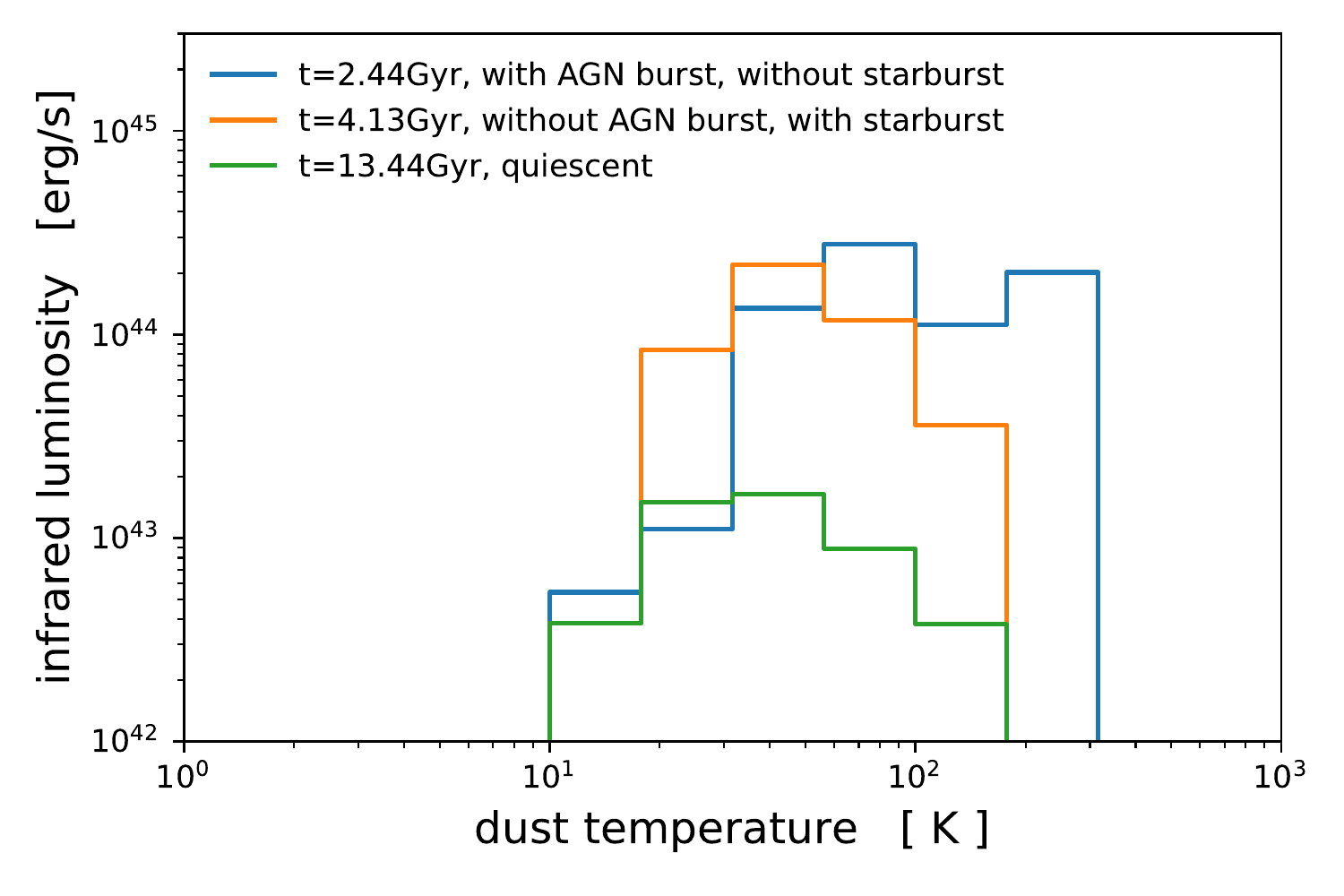}
\caption{Histogram of infrared luminosity in dust temperature bins. The vertical axis is the total instantaneous infrared luminosity integrated over the galaxy within given dust temperature bin.}
\label{fig:dust-temperature-histogram}
\end{figure}

\begin{figure}[htb]
\centering
\includegraphics[width=0.475\textwidth]{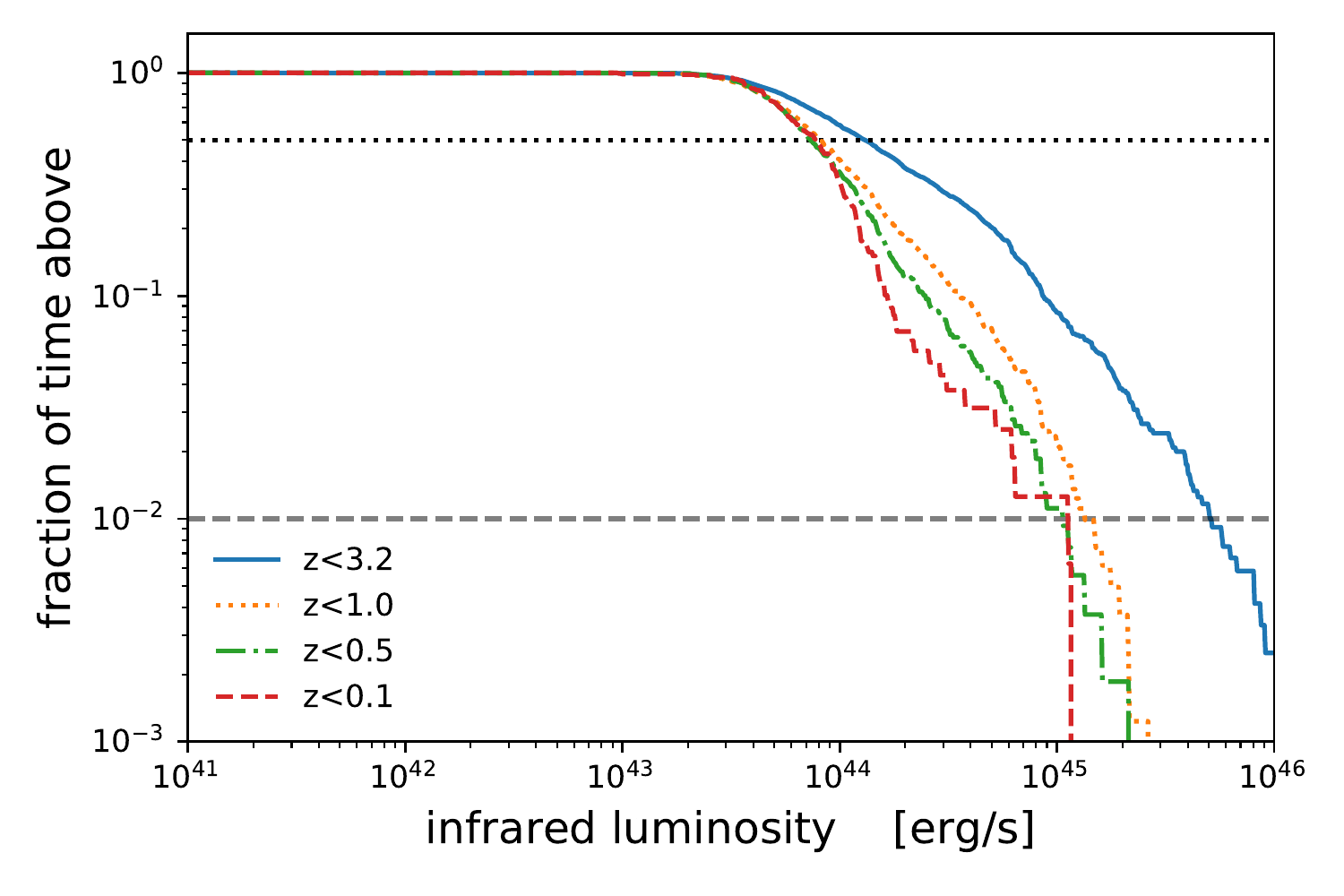}
\caption{Duty cycle of the infrared luminosity from dust, i.e., the fraction of time when the infrared luminosity is above given values. 
The horizontal dotted line is at 50\% level and dashed at 1\%. }
\label{fig:dust-IR-duty-cycle}
\end{figure}

\begin{figure*}[htb]
\centering
\includegraphics[width=0.475\textwidth]{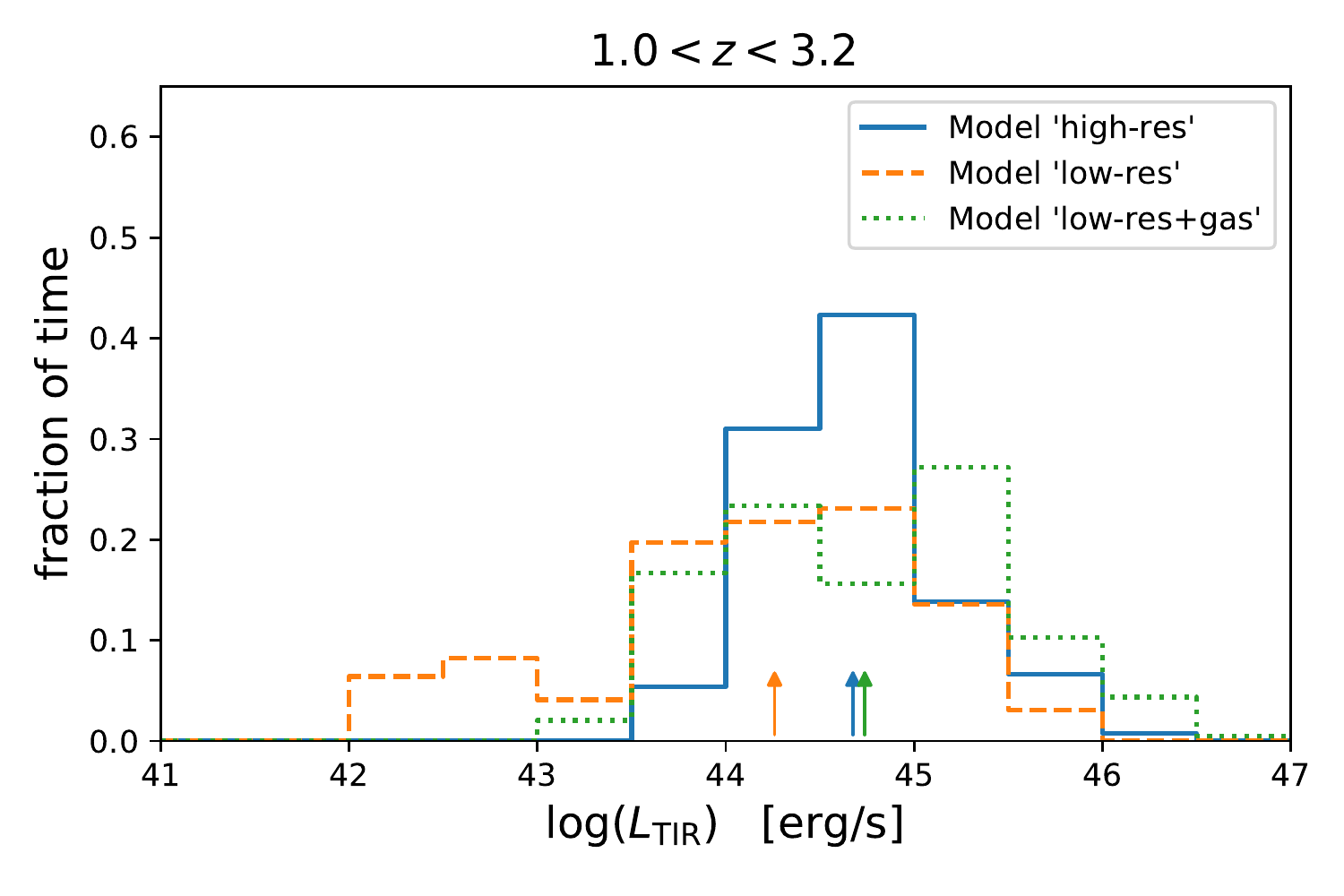}
\includegraphics[width=0.475\textwidth]{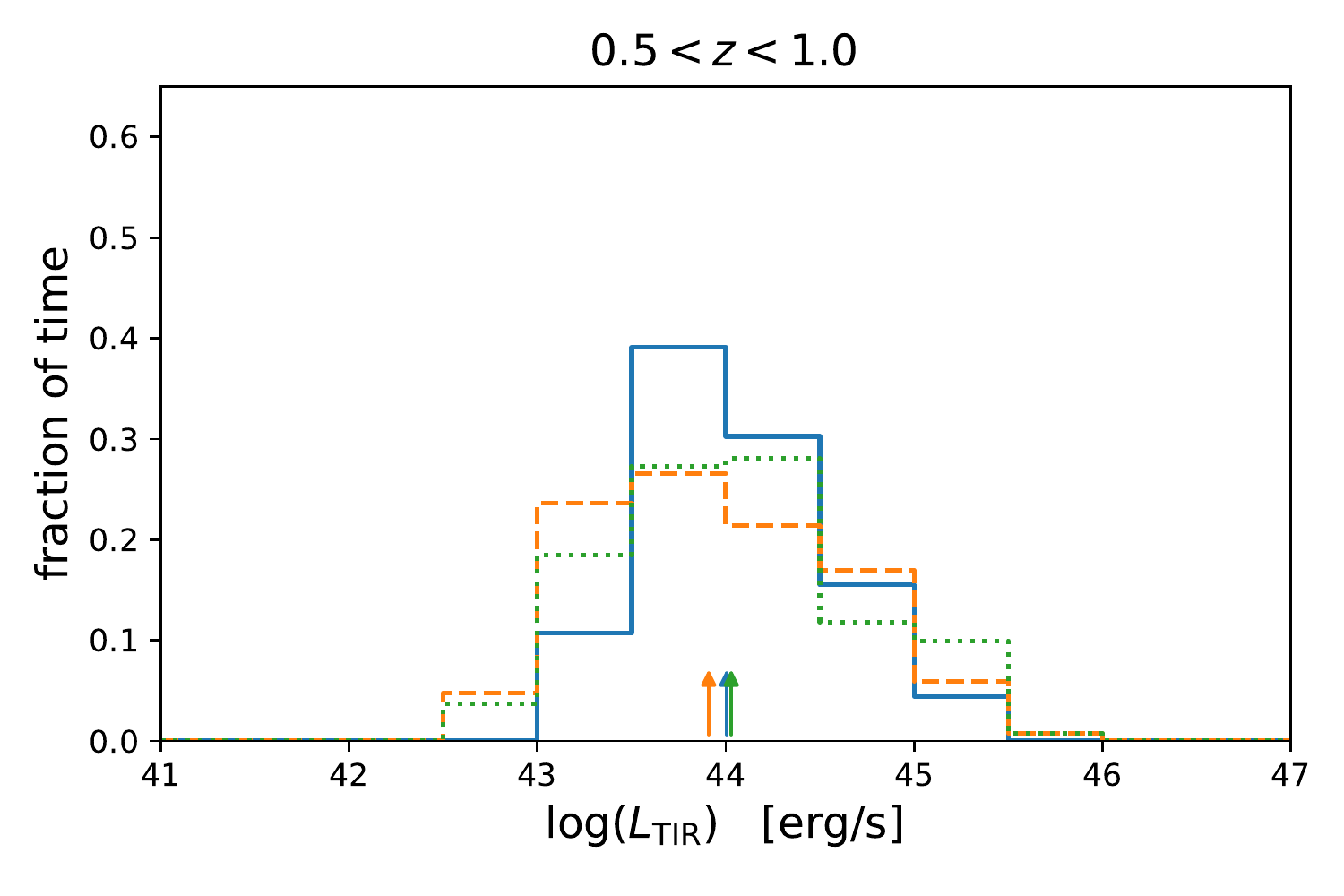}
\includegraphics[width=0.475\textwidth]{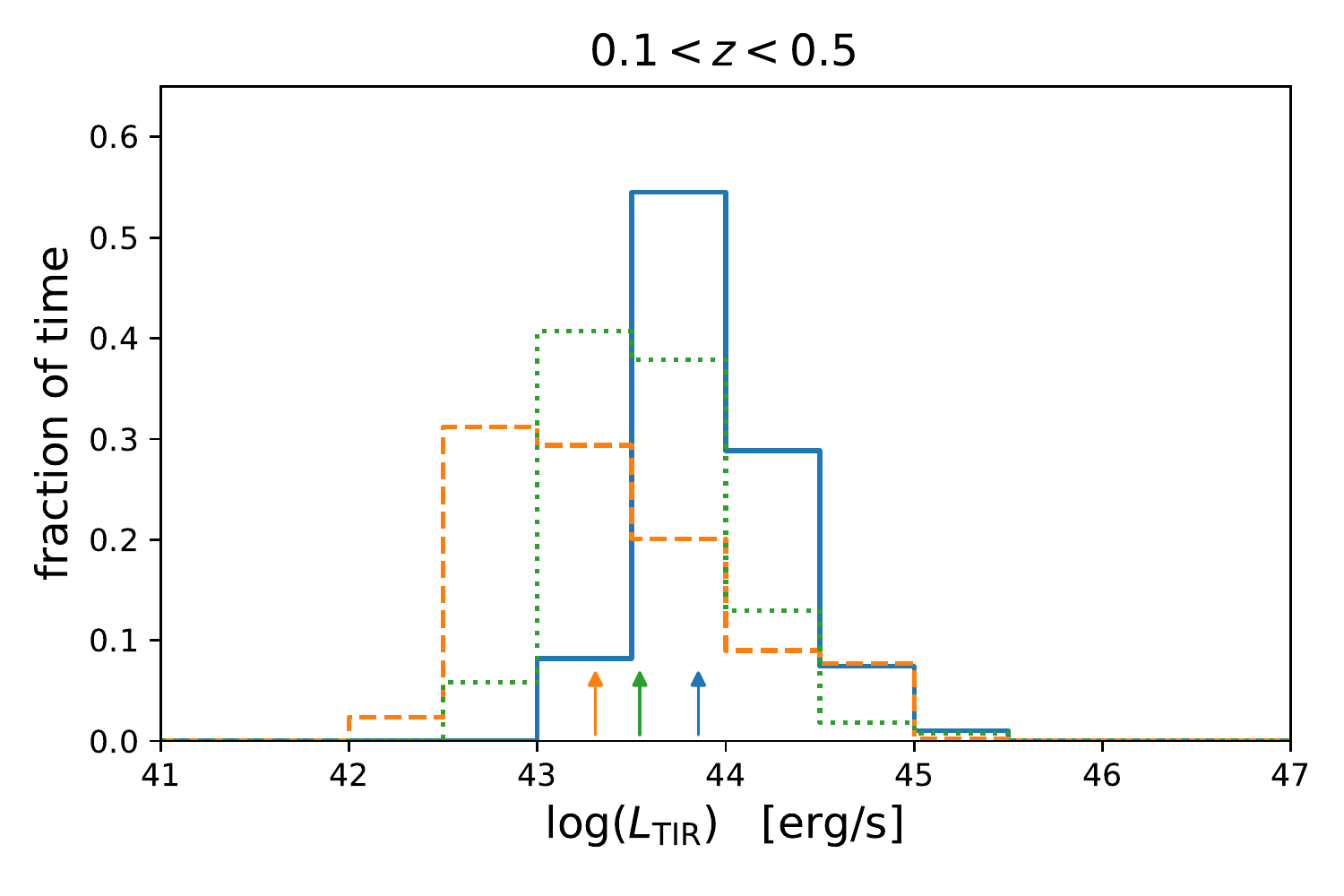}
\includegraphics[width=0.475\textwidth]{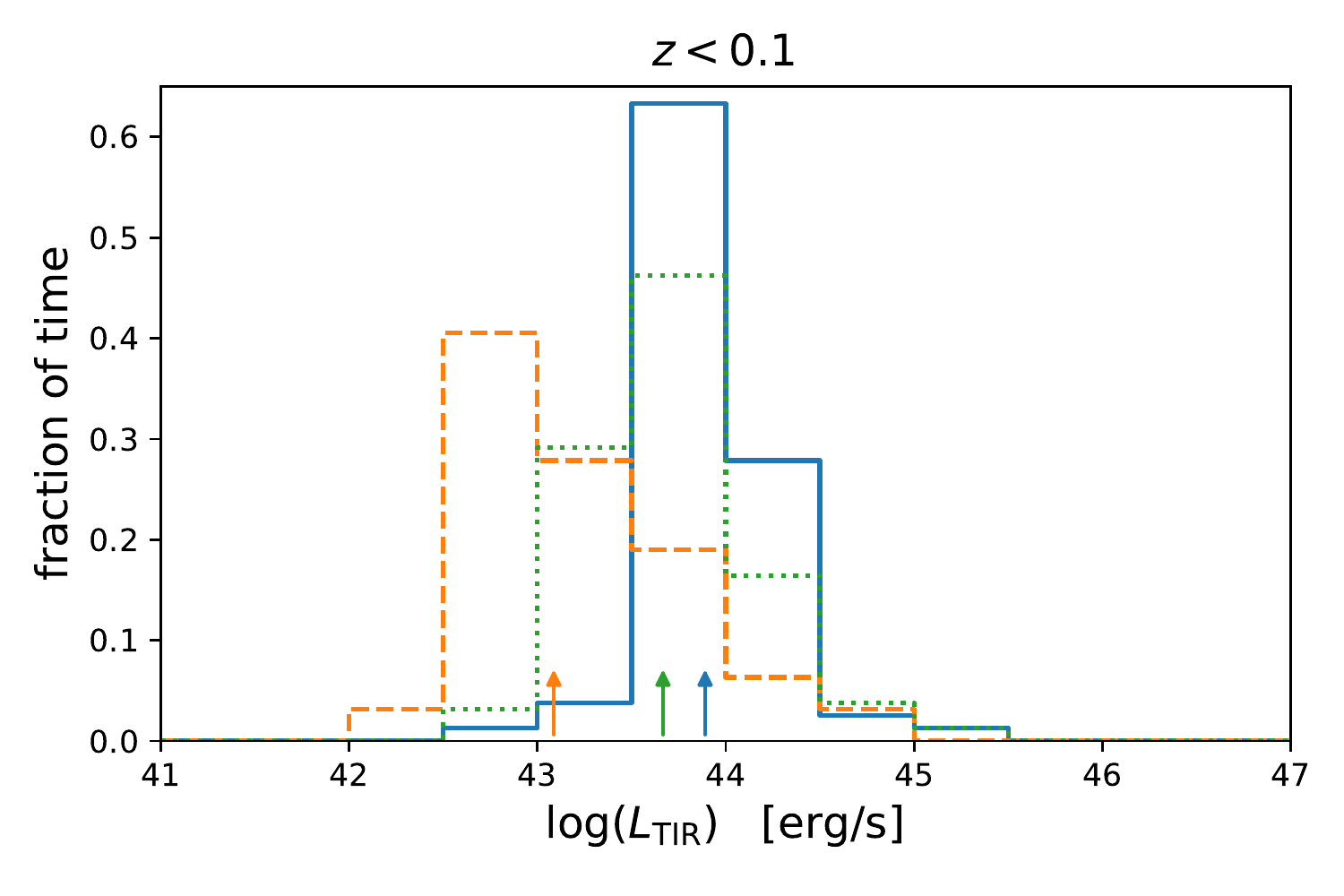}
\caption{histogram of Infrared luminosity from dust. The vertical arrows indicate the median values of the infrared luminosity.}
\label{fig:dust-IR-emission-histogram}
\end{figure*}

In Figure \ref{fig:dust-IR-emissivity}, we plot the spatial distribution of the dust emission during the AGN outburst 
\textcolor{black}{at $t=2.44$ Gyr}.
 It is shown that the dust emission during the outbursts 
is mostly due to the AGN irradiation of the innermost region 
and the surface of the disk where the optical depth to the central AGN is not obscured.
As other evidence, in the bottom panel of Figure \ref{fig:IR-half-radius} 
we plot the scatter of the half-light radius of the dust infrared emission vs the total infrared luminosity.  
We can see that the half-light radius decreases moderately when the infrared luminosity increases,
with typical values of approximately \textcolor{black}{30 parsec at $2\times10^{44}$ erg/s}.
In the top panel of Figure \ref{fig:IR-half-radius}, 
we plot the scatter of the infrared emission weighted dust temperature vs the total infrared luminosity, 
where a tight correlation is found.
While most of the gas mass is in the X-ray emitting, extended hot component,
most of the infrared radiation is emitted by the highly concentrated cold component.

In Figure \ref{fig:dust-temperature-histogram}, we present the histogram of infrared luminosity in dust temperature bins, 
in which the vertical axis is the total instantaneous infrared luminosity integrated over the galaxy within given dust temperature bin. 
The colored lines are the results from three representative time snapshots, i.e., 
\textcolor{black}{$t=2.44$ Gyr} during an AGN burst while without starburst; 
\textcolor{black}{$t=4.13$ Gyr} when without AGN burst while with starburst; and
\textcolor{black}{$t=9.39$ Gyr} when the galaxy is quiescent, respectively 
(see also the samples in Figure \ref{fig:IR-half-radius} for more details). 
We can see the dust temperature peaks around $\sim 10^2$ K, 
and it is systematically higher when the infrared luminosity increases, 
especially during the AGN outbursts when the dust heating is dominated by the AGN irradiation.
The dust temperature can reach \textcolor{black}{$\sim500$ K} in the innermost region near to the central AGN,
while the typical dust temperature in the circumnuclear disk is \textcolor{black}{$\sim30$ K}.

In Figure \ref{fig:dust-IR-duty-cycle} we present the duty cycle of the infrared luminosity from dust, 
i.e., the fraction of time when the infrared luminosity is above given values.
It is shown that most of time the infrared luminosity is \textcolor{black}{$>3\times10^{43}$ erg/s}, and it systematically
decreases when the redshift decreases. 
The median value is \textcolor{black}{$\sim2\times10^{44}$ erg/s} (the blue line), 
or \textcolor{black}{$\sim1/2$} of that at late times (the red line).

In Figure \ref{fig:dust-IR-emission-histogram}, we plot the histograms of the dust infrared emission.
It is shown that there is a clear red-shift dependency
of the dust infrared luminosity, i.e., the infrared luminosity decreases systematically
as the redshift decreases. Again, it is due to the AGN irradiation (which contributes the most energy for dust emission),
as there is more AGN activity at high redshift (see also Figures \ref{fig:sim-overview} and \ref{fig:duty-cycle}). 
In all of the histograms, we can also see a cut-off around the infrared luminosity \textcolor{black}{$\sim3\times10^{43}$ erg/s}.
We find in the simulation that the collisions between dust grains and gaseous medium always contribute
significant infrared emission, no matter whether there are AGN burst and/or star formation,
though the latter may boost the dust collisional heating via feedback processes. 
As a consequence, there is a ``floor'' luminosity due to the collisional heating 
(\textcolor{black}{$\sim10^{43}$ erg/s} in our simulated galaxy; see the green line in the top panel of  Figure \ref{fig:dust-IR-emission}),
which results in the cut-off in the histogram. 
The stellar irradiation by the embedded new stars also contributes a significant fraction of energy for the dust infrared emission
(see the orange line in the top panel of Figure \ref{fig:dust-IR-emission}). 

\begin{table*}[ht] \label{tab:experiment-list}
\caption{Simulation outputs of the model galaxy $^*$}
\label{tab:experiment-list}
\begin{center}
\footnotesize
\begin{tabular}{ccccccccccccc}
\hline\hline
\multirow{2}{*}{ Model } &\multicolumn{3}{c}{AGN duty cycle$~~^a$} & {$M_{\rm BH}$}$^b$ & ${\rm SF}$$^c$ & $\left<{\rm SFR}\right>$$^{d}$ 
& $\left<L_{\rm X}\right>$$^{e **}$ & $\left<R_{\rm 0.5}^{X}\right>$$^f$ & $\left<L_{\rm TIR}\right>$$^g$ & $\left<R_{\rm 0.5}^{IR}\right>$$^h$ & $\left<M^{\rm cold}_{\rm gas}\right>$$^i$ & $R_{\rm disk}$$^j$  \\ 
\cline{2-4}
{} &{$f_{l>0.1}$}&{$l_{\rm median}^{\rm energy}$}&{$l_{\rm median}^{\rm time}$} &$10^9 M_\odot$&$10^9 M_\odot$&$M_\odot$/yr&$10^{42}$erg/s&kpc&$10^{43}$erg/s&kpc&$10^9 M_\odot$& kpc\\
\hline
high-res & 7.62\% & $0.011$ & $0.00094$ & $6.72$   & $32.9$ & 1.24& $0.07$ & 14.5 & 8.23 & 0.03 & 0.97 & 1.53\\
low-res & 9.59\% & $0.025$ & $0.00042$ & $8.32$   & $9.84$ & 0.15 & $0.10$ & 31.6 &  1.60 & 0.41 & 2.05 & 1.52\\
low-res+gas & 37.0\% & $0.039$ & $0.00026$ & $25.6$   & $26.4$ & 0.33 & $0.33$ & 19.5 & 5.39 & 0.40 & 3.06 &1.67\\
\hline \hline\\
\end{tabular}
\end{center}

\footnotesize
* \textcolor{black}{From left to right, the columns in the table above present the simulation results: 
a) AGN duty cycle, including the percentage of the cumulative radiation energy that the AGN emits when its Eddington ratio $l\equiv L_{\rm AGN}/L_{\rm Edd}$ is above 0.1 (i.e., $f_{l>0.1}$), 
                                              the median Eddington ratio above which the AGN emits half of its total energy (i.e., $l_{\rm median}^{\rm energy}$), and
                                              the median Eddington ratio above which the AGN spends half of the simulated cosmological timespan (i.e., $l_{\rm median}^{\rm time}$);
b) the final black hole mass $M_{\rm BH}$ (the initial black hole is $0.61\times10^9M_\odot$ in all simulations); 
c) total star formation;  d) averaged star formation rate ($\left<{\rm SFR}\right>$);
e) averaged X-ray luminosity from the hot ISM ($\left<L_{\rm X}\right>$); f) averaged half-light radius of the X-ray emission ($\left<R_{\rm 0.5, X}\right>$);
g) averaged infrared luminosity from dust ($\left<L_{\rm TIR}\right>$); h) averaged half-light radius of the infrared emission ($\left<R_{\rm 0.5, IR}\right>$);
i) averaged mass of the cold gas with temperature lower than 100 K ($\left<M^{\rm cold}_{\rm gas}\right>$);
j) size of the circumnuclear disk at the end of the simulations ($R_{\rm disk}$).} \\
** \textcolor{black}{All the averages are made using the simulation data in the last 1 Gyr.} \\
                             
\end{table*}

\subsection{Dependence on Initial Condition and Resolution} \label{sec:resolution-study}

In this subsection we study the dependence of our results on the initial conditions and spatial resolution. 

In the fiducial model we have adopted the previously standard assumption that, after formation,
elliptical galaxies have a low gas fraction, and so for simplicity we took a gas free initial state 
with all subsequent gas content due to cosmological infall or mass loss from evolving stars.
This fiducial model is model ``high-res'' in Table \ref{tab:experiment-list} and it has an inner boundary of 2.5 parsec. 
To determine the importance of our high spatial resolution, we include as model ``low-res'', 
a simulation which is identical except for an inner boundary of 25 parsec,
which is still small compared to the resolution of most cosmological simulations.

We see some dramatic changes in the lower resolution model. 
The infrared luminosity is much higher and the dusty disk is much smaller in the high resolution model,
and the new stars are much more numerous and concentrated when resolution is increased.
The black hole final mass is less as a consequence, 
since the gas flowing in towards the center can be turned into stars before reaching the central black hole.

The implication is strong that a still higher spatial resolution
(currently beyond technical reach) would further enhance these trends. 
It is possible however that heating from the central black hole would limit star formation 
within, as we see in Figure \ref{fig:hydro-properties}, $r<10$ parsec.

The other variation on our fiducial model is based on starting the simulation with a gas/star ratio of 30\%, 
close to the value seen in recent ALMA/SCUAB-2 survey of submillimeter galaxies \citep{dudzeviciute_alma_2019}. 
That work (cf. their Fig 11d) indicates a gas/star ratio of roughly 20\% - 40\% at redshift $z=2$.
To save computer time, we ran the simulation with 30\% initial gas fraction which was identical 
to the low resolution model ``low-res'', labelling the high initial gas model ``low-res+gas'' in Table \ref{tab:experiment-list}. 
We see several important changes. 
First, it is much more ``bursty'' with  \textcolor{black}{37\%} of the AGN radiation energy emitted above $L/L_{\rm Edd}=0.1$.
\textcolor{black}{As expected,} there is more star formation, more infrared emission and a higher X-ray luminosity.

Figure \ref{fig:dust-IR-emission-histogram} shows the distribution of infrared luminosities for the three simulations
as a function of redshift range. Both higher resolution and more initial gas push up the infrared luminosities to $10^{44}$ erg/s
at late times and $10^{45}$ erg/s at redshift above $z=2$.

A model with both higher resolution and more initial gas is being studied and will be reported on in a subsequent communication.


\section{Comparison to Observations} \label{sec:compare-to-obs}

While earlier 1-D models incorporating the grain physics 
have already been compared to observations in \cite{hensley_grain_2014} 
\textcolor{black}{(previous attempts to model the dust emission and its SED can be also found in \citealt{schurer_modelling_2009})}, 
the current 2-D models allow for a more detailed comparison across redshift and galaxy properties. 
Existence of cold gas rotating disks in the more realistic 2D models greatly enhances infrared (IR) emission.

\subsection{High Redshift}

Extending to high redshift, \cite{Schreiber_dust_2018} fit models to stacked SEDs from deep CANDELS fields, 
including stacked Herschel and ALMA data. Although their model templates applied to the observations 
do not include emission from a dusty AGN torus, this should have little impact on their derived $L_{\rm TIR}$ and stellar masses.  
For their highest stellar mass bin ($11<\log M_\star<11.5$) both data and models 
suggest stacked-average $L_{\rm TIR}$ $\sim$ 3, 7, 17, 29, 
and 51 $\times 10^{44}$ erg s$^{-1}$ at z $\sim$ 0.5, 1, 1.5, 2.2, and 3, respectively. 
These L$_{\rm TIR}$ values are in excellent agreement with our predictions (cf. Figure \ref{fig:dust-IR-emission-histogram}).

Such high-luminosity massive galaxies are likely to evolve into the quiescent massive galaxies observed in the local universe.  
For example, using ALMA, \cite{wang_dominant_2019} recently discovered a population of very massive, optically-faint, 
IR-luminous galaxies at z$\sim$4 with average L$_{\rm TIR}$ = 40$\times 10^{44}$ erg s$^{-1}$. 
This population has a  clustering/bias factor consistent with the population 
expected to be the progenitors of massive galaxies in groups and clusters in the local Universe, 
such as the one modeled here.

Results from \cite{Schreiber_dust_2018} suggest that at $z>$1, 
massive galaxies dominate the luminous end of the TIR luminosity function.  
Our model suggests high IR variability with a broad, flattened distribution over log L$_{\rm TIR}$.  
Measurements of the TIR (or FIR) luminosity function 
(e.g. \citealt{gruppioni_herschel_2013} and \citealt{koprowski_evolving_2017}) 
are relatively consistent with the range and the shape of the distribution we infer. 
However these LFs are themselves discrepant on the bright end, 
possibly due to differing selection effects with respect to AGN or galaxies 
with a warmer dust component (see, e.g. \citealt{gruppioni_existence_2019}).


\subsection{Low Redshift} \label{sec:IR-obs-low-redshift}

At low redshifts, for $z<0.5$, our model predicts a median infrared luminosity \textcolor{black}{$\sim 1.2\times 10^{44}$ erg s$^{-1}$},  
and 90\% of the time, \textcolor{black}{L$_{\rm TIR}<4\times10^{44}$ erg s$^{-1}$} (cf. Figure \ref{fig:dust-IR-duty-cycle}).  
While luminosities fall just below the lower bound for the typical definition for Luminous Infrared Galaxies (LIRGs), 
these luminosities are nevertheless comparable or even higher than those of less massive star-forming galaxies. 
There is evidence that some relatively quiescent massive galaxies have TIR luminosities in this range.

\cite{Andreani_bivariate_2018} measured the bivariate luminosity and mass functions 
from a local sample from the Herschel Reference Survey.  
For galaxies with high stellar masses (e.g. log M$_\star$ $>$ 11.5), 
their bivariate function implies a broad range of TIR luminosities centered on 1-2$\times 10^{43}$ erg s$^{-1}$.  
Their analysis only includes galaxies classified as late-types, 
but the massive galaxies included in their analysis are likely to contain a significant spheroid component.

Surveys of IR emission from dust in early-types in the nearby Universe 
typically contain only a handful galaxies with high stellar masses ($\log M_\star > 11$). 
\cite{Amblard_star_2014} extract physical properties from 221 early type galaxies of which  $\sim$50 have high stellar masses. 
The median $L_{\rm dust}$ of this subsample is $10^{42}$ erg s$^{-1}$, with 5 (10\%) above $10^{43}$ erg s$^{-1}$. 
Infrared AKARI observations of 260 early-type galaxies in the Atlas3D sample 
(\citealt{Kokusho_star_2017}; \citealt{Kokusho_dust_2019}) do include several 10s of high mass galaxies, 
with a spread of $L_{\rm TIR}$ in the $10^{42}$ to 10$^{43}$ erg s$^{-1}$ range, and occasionally higher.  
Interestingly, they note much wider scatter in warm dust luminosities (at fixed stellar mass), 
and conclude that this is likely due to variation in AGN activity across the sample.  
Galaxies with dusty nuclei and/or post-starburst signatures also show high IR luminosities.  
For example, in a sample of E+A galaxies \citealt{Smercina_after_2018} 
measure L$_{\rm TIR}$ $\sim 10^{43.5}$ and 10$^{44}$ erg s$^{-1}$ for two galaxies with high stellar mass.


\section{Conclusions and Discussions} \label{sec:conclusions}

We have developed the \texttt{MACER} (Massive AGN Controlled Ellipticals Resolved) code 
for exploring the evolution of massive elliptical galaxies at a high spatial resolution down to and within the fiducial Bondi radius. 
Though it is still a two-dimensional code, the high time and space resolution enables us to study the coevolution 
between the supermassive black holes and their host galaxies self-consistently, 
which is difficult for large-scale cosmological simulations.
In our hydrodynamical simulations, we have included a relatively complete set of stellar physics,  
galaxy dynamics, and AGN feedback processes. 
In the presence of significant rotation, 
circumnuclear disks form in the center of the elliptical galaxies, 
which is a natural consequence of standard cooling flows, but the mass inflow is obstructed 
because of its angular momentum barrier \citep{gan_macer_2019}. 
The circumnuclear disk is found to be cold and dense \textcolor{black}{with typical mass of $10^9M_\odot$},
which makes it an ideal site for star formation when it becomes self-graviting. 
The latter also permits angular momentum transfer 
and thus allows mass inflow onto the central supermassive black hole (after some time lag). 
Therefore, one can always expect a near coincidence between star formation and AGN activities 
(with rates even exceeding those of spiral galaxies), 
though we found in the simulations that in most of their lifetime the simulated galaxies 
are stereotypical ``quiescent" elliptical galaxies. 

In our previous paper \citep{gan_adding_2019}, we added a suite of chemical abundances into the \texttt{MACER} code 
to track the metal enrichment due to the passive stellar evolution of AGB stars and supernovae of type Ia and II. 
However, we did not consider dust grains, which are known to be important in depleting metals. 
Especially in the circumnuclear disk, the cold and dense environment favors the dust grain growth. 
Therefore, a dusty disk can be expected, and it can also be important in absorbing the starlight within the disk 
and in obscuring the AGN radiation from the galaxy center. 
\textcolor{black}{Such dusty disks are by now well observed.}
The computed infrared emission from the dust may also provide a diagnosis to test our model observationally. 

\textcolor{black}{
Following \citet{hensley_grain_2014}, we implement the following grain physics into our code 
(assuming dust grains co-move with the gaseous ISM): 
(a) dust grains made and injected in the passive stellar evolution. We assume that most of the metals in the stellar ejecta are in dust grains; 
(b) dust grain growth due to collision and sticking; 
(c) dust grain destruction due to thermal sputtering; 
(d) dust cooling of hot gas via inelastic collisions; 
(e) radiation pressure on dust grains. 
}

The representative galaxy model adopted for the simulation has a  stellar mass of $6.1\times10^{11}M_\odot$, 
a central stellar velocity dispersion of $\sim260$ km/s, and  ellipticity 0.37.  
We studied the time evolution of the dust, including its spatial distribution, its depletion of metals, 
its absorption of the starlight within the circumnuclear disk, 
its obscuration of the central AGN radiation, and finally its infrared emission. 
We find that:
\begin{itemize}
\item 
In \textcolor{black}{more than half ($\sim55\%$)} of its lifetime, the simulated galaxy is a stereotypical ``quiescent" elliptical galaxy 
with little star formation (\textcolor{black}{$\lesssim2~M_\odot/{\rm yr}$}). 
However, during the central outbursts, the star formation rate can be up to \textcolor{black}{$>250~M_\odot/{\rm yr}$}.
All star formation occurs within the circumnuclear disk, which is optically thick. 
The size of the disk is $\sim 1$ kpc, while the half-mass radius of the resulting stars is of order \textcolor{black}{20 parsec.}

\item 
It is dusty in the outer disk \textcolor{black}{($r\sim1$ kpc)} where most of the metals are in dust grains, 
while in the inner disk \textcolor{black}{($r\lesssim20$ parsec)} most of the dust grains are destroyed by 
thermal sputtering in the hot, AGN-irradiated gas.

\item 
The dusty disk is optically thick to both the starlight within the disk and the radiation from the central AGN. 
The AGN is expected to be obscured behind the circumnuclear disk, 
having a covering factor of $\sim0.2$.

\item
The total infrared radiation from the dust is greater than the X-rays from the hot gas. 
The median infrared luminosity from the dust is \textcolor{black}{$\sim2\times10^{44}$ erg/s}, 
and it can reach up to $\sim10^{46}$ erg/s during outbursts. 
Most of the dust infrared emission is due to the AGN irradiation, 
and it is located at the innermost region and on the surface of the dusty disk 
where the dust optical depth is moderate to the AGN radiation. 
The half-light radius of the infrared emitting region is \textcolor{black}{$\sim30$ parsec} during bursts.
\end{itemize}

One test of our modeling would come from the measurement of  half-light radii ($R_e$). 
In the X-ray regions, typical sizes are computed to be $\sim100$ kpc.
In the optical (by assumption), the half-light radius is $\sim10$ kpc. 
It is striking that our computed infrared sizes are typically much smaller --- 30 parsec ---
and this prediction should be testable by JWST.

\textcolor{black}{
As future work, we also plan to extend our simulations into three dimensions. 
Our current treatments for some important physics processes are limited by the two-dimensional settings.
For example, the assumption of axial symmetry breaks in the Toomre instability, 
which is intrinsically three-dimensional --- it is extremely important to both AGN feeding and star formation.
It is also true for supernova explosions, in our current two-dimensional settings, 
we have to average the SN-ISM energy coupling in the axial direction, 
i.e., the SN heating is averaged over rings, rather than disposing its energy locally, 
where the latter would induce more anisotropy to the X-ray emitting ISM as observed. }


\section*{Acknowledgement}
We thank David Spergel and the Center for Computational Astrophysics for
	 generous support of this work and also James Stone and Princeton University
	 for their generous support as well.
We thank Jeremy Goodman, James Stone, Kengo Tomida, Pieter van Dokkum, 
	 Nadia Zakamska, Takayuki Saitoh,Tuguldur Sukhboldfor, Charlie Conroy,
	 Feng Yuan and Doosoo Yoon 
	 for useful discussions.
We thank Gregory S. Novak for sharing the first 2D version of
	 the \texttt{MACER} code in 2011, which was using \texttt{ZEUSMP/1.5}.
We acknowledge computing resources from Columbia University's 
 	 Shared Research Computing Facility project,
  	 which is supported by NIH Research Facility Improvement 
 	 Grant 1G20RR030893-01, and associated funds from 
 	 the New York State Empire State Development, Division of Science Technology 
 	 and Innovation (NYSTAR) Contract C090171, both awarded April 15, 2010. 
We are also pleased to acknowledge that the work reported on in this paper 
	was substantially performed using the Princeton Research Computing resources at Princeton University 
	which is consortium of groups including the Princeton Institute for Computational Science and Engineering 
	and the Princeton University Office of Information Technology's Research Computing department.

\bibliography{ms_macer_dust}  
\end{document}